\documentclass[12pt,a4paper]{article}
\usepackage{amsmath,amsfonts,epsfig}
\usepackage{amssymb}
\usepackage{mathrsfs}
\usepackage{hyperref}
\newcommand{\beq}{\begin{eqnarray}}
\newcommand{\eeq}{\end{eqnarray}}

\newcommand{\be}{\begin{equation}}
\newcommand{\ee}{\end{equation}}

\newcommand{\bm}{\begin{multline}}
\newcommand{\fm}{\end{multline}}

\begin{document}
\numberwithin{equation}{section}
\setlength{\unitlength}{.8mm}

\begin{titlepage} 
\vspace*{0.5cm}
\begin{center}
{\Large\bf Lattice approach to finite volume form-factors of the Massive Thirring/Sine-Gordon model}
\end{center}
\vspace{1.5cm}
\begin{center}
{\large \'Arp\'ad Heged\H us}
\end{center}
\bigskip

\vspace{0.1cm}

\begin{center}
Wigner Research Centre for Physics,\\
H-1525 Budapest 114, P.O.B. 49, Hungary\\ 
\end{center}
\vspace{1.5cm}
\begin{abstract}

\end{abstract}
In this paper we demonstrate, that the light-cone lattice approach for the Massive-Thirring (sine-Gordon)
model, through the quantum inverse scattering method, admits an appropriate framework for computing the
finite volume form-factors of local operators of the model.
In this work we compute the finite volume diagonal matrix elements of the $U(1)$ conserved current in the pure
 soliton sector of the theory.
Based on the systematic large volume expansion of our results, we conjecture an exact expression for the 
finite volume expectation values of local operators in pure soliton states. At large volume in leading order 
these expectation values have the same form as in purely elastic scattering theories, but exponentially small 
corrections differ from previous Thermodynamic Bethe Ansatz conjectures of purely elastic scattering theories. 

\end{titlepage}

\section{Introduction}

The computation of finite volume matrix elements of local operators is an important problem in integrable quantum field theories. 
These form-factors play important role in the determination of heavy-heavy-light 3-point functions 
in the planar $AdS_5/CFT_4$ correspondence \cite{Bjhhl}, and they are fundamental ingredients of the form-factor perturbation theory \cite{DMS96}.

In \cite{ddvlc} the Massive Thirring (MT) model was formulated as the continuum limit of an inhomogeneous 6-vertex model with
appropriately chosen alternating inhomogeneities. This integrable lattice regularization allowed one to compute the
finite volume spectrum of the theory by solving a set of nonlinear-integral equations (NLIE) \cite{KP1}-\cite{FRT3}. Due to the 
bosonization link between the Massive Thirring and sine-Gordon models \cite{s-coleman,klassme}, this method gave access to the finite volume 
spectrum of the sine-Gordon (SG) model as well. The NLIE description of the finite volume spectrum was checked against direct field 
theoretical methods such as the Truncated Conformal Space Approach (TCSA) as well \cite{FRT2}.

Nevertheless, the integrable lattice regularization of \cite{ddvlc} gives access to compute matrix elements of local operators 
of the MT model and of their bosonized counterparts in the SG model. The general framework for these computations is the 
Quantum Inverse Scattering Method (QISM) \cite{FST79}. In the past decades a remarkable amount of progress has been achieved 
in the computation of form-factors and correlation functions of local spin operators on the lattice \cite{book}-\cite{LukTer}. One of the 
most important discovery was that local spin operators can be expressed in terms of the elements of the Yang-Baxter 
algebra in an elegant way \cite{KMT99}. This made it possible to compute the matrix elements of local spin operators by
using only the Yang-Baxter algebra. 

 Relying on the light-cone lattice regularization of \cite{ddvlc}, in this paper our purpose is to 
compute finite volume form-factors of local operators in the MT/SG theories.
The lattice Fermi fields of the regularized MT model are related to the spin operators by 
a Jordan-Wigner transformation. This is why the results of 
the QISM for spin variables are directly applicable to our model. Nevertheless, due to renormalization 
effects{\footnote{Here we think to normal ordering, renormalization constants and operator mixing.}} the connection 
between lattice fields and the fields of the continuum theory can be very non-trivial.
Because of these subtleties in this paper we restrict ourselves to operators which are related to the $U(1)$ 
symmetry of the model. Since the $U(1)$ symmetry is present in both the lattice and the continuum theories, it makes 
easier to make a connection between the lattice and the continuum fields. The principle is that conserved quantities of 
the regularized theory are mapped to conserved quantities of the continuum model. In this manner we can identify the two 
components of the conserved $U(1)$ current{\footnote{The corresponding conserved quantity is the toplogical charge in the SG model.}} 
of the continuum theory as $J_0(x_n)\sim \frac{\sigma^z_{2n}+\sigma^z_{2n-1}}{2},\\ \quad J_1(x_n)\sim \frac{\sigma^z_{2n}-\sigma^z_{2n-1}}{2}$.

Using the QISM techniques the diagonal matrix elements of $J_\mu$ can be computed on the lattice and the continuum limit 
can be taken as well. The final results can be expressed in terms of the counting-function of the theory, which satisfies
 a set of NLIEs \cite{FRT1}-\cite{FRT3}, which we will refer to as DDV equations. For the sake of simplicity, 
in our actual computations we restricted ourselves to the pure 
soliton sector{\footnote{In lattice terminology: we restrict ourselves to pure hole states over the antiferromagnetic vacuum.}} of the theory, but the computations could be extended without any serious difficulties to other excited states of the model, as well. 

For $J_0$ we got the expected and quite trivial result, that the expectation value is equal to the topological charge of the
state divided by the volume. For $J_1$ the result is not so trivial. There the expectation value can be expressed by the 
solution of a linear integral equation, whose kernel depend on the counting-function of the sandwiching state. These  equations can be solved analytically in the context of a systematic large volume expansion.

It turns out, that in accordance with \cite{Palmai13}, 
in the pure soliton sector, at large volume in leading order the diagonal form-factors of $J_\mu$ can be expressed 
in terms of the so-called connected-form factors of the operator in exactly the same way as in purely elastic scattering theories \cite{Pozsg13,PST14}.
Nevertheless the exponentially small in volume corrections differ from the TBA conjectures \cite{LM99,saleur,Pozsg11,Pozsg13,PST14} of purely elastic scattering 
theories. The difference arises in the form of the so-called dressed-form factors,
 which in our case are functionals of the counting-function of the sandwiching state and the connected-form factors of $J_\mu$ (\ref{JDdress}).

Based on previous experiences in diagonally scattering theories, we conjecture that in the pure soliton sector, our final 
 formula (\ref{JDdress}) for the dressed form-factors hold for any operator, provided the connected form-factors 
of the operator under consideration is substituted into (\ref{JDdress}).

The organization of the paper is as follows. 
In section 2. we summarize the light-cone lattice approach to the MT model and determine the lattice counterparts of the
$U(1)$ conserved current. The NLIE governing the finite volume spectrum of the model is also reviewed in this section.
In section 3. we provide the integrable QISM formulation of the model.  In section 4. the diagonal matrix elements of 
the operator $\sigma_n^z$ are computed on the lattice. 
The continuum equations, their solution and the correct identification between the lattice and continuum fields are presented in section 5.
The systematic large volume expansion and the determination of dressed form factors can be found in section 6.
 Our summary and outlook can be found in section 7.
The paper includes a short appendix containing some Fourier-transforms being necessary for the computations.

\section{Light-cone approach to the Massive-Thirring/sine-Gordon models}

The continuum models we consider in this paper are the sine-Gordon theory,
\begin{equation}
\label{sG_Lagrangian}
{\cal L}_{SG}= \displaystyle\frac{1}{2}\partial _{\nu }\Phi \partial ^{\nu }\Phi +\displaystyle\frac{\mu ^{2}}{\beta ^{2}}:\cos \left( \beta \Phi \right) :\,  \qquad 0<\beta^2<8 \pi,
\end{equation}
and the massive Thirring model:
\begin{equation}
\label{mTh_Lagrangian}
{\cal L}_{MT}= \bar{\Psi }(i\gamma _{\nu }\partial ^{\nu }+m_{0})\Psi -\displaystyle\frac{g}{2}\bar{\Psi }\gamma^{\nu }\Psi \bar{\Psi }\gamma _{\nu }\Psi \,,
\end{equation}
where we use chiral representation for the fermions $\{\gamma^\mu,\gamma^\nu\}=2 \eta^{\mu \nu}$: 
$$ \Psi=\begin{pmatrix} \psi_L \\ \psi_R \end{pmatrix}, \quad \gamma^0=\begin{pmatrix} 0 & 1 \\ 1 & 0 \end{pmatrix}, \quad \gamma^1=\begin{pmatrix} 0 & 1 \\ -1 & 0 \end{pmatrix}, \qquad
\gamma^5=\gamma^0 \gamma^1=-\eta=\begin{pmatrix} -1 & 0 \\ 0 & 1 \end{pmatrix}.$$
By bosonization techniques, it was shown \cite{s-coleman} that the two models can be mapped into each other provided their coupling constants satisfy the relation: 
\begin{equation} \label{gbeta}
1+\frac{g}{4 \pi}=\frac{4 \pi}{\beta^2}.
\end{equation}
There is a subtle point in the equivalence of the two theories \cite{klassme}, namely they are equivalent only in the even topological charge sector of 
their Hilbert-spaces and they differ in the odd topological charge sector.

\begin{figure}[htb]
\begin{flushleft}
\hskip 15mm
\leavevmode
\epsfxsize=120mm
\epsfbox{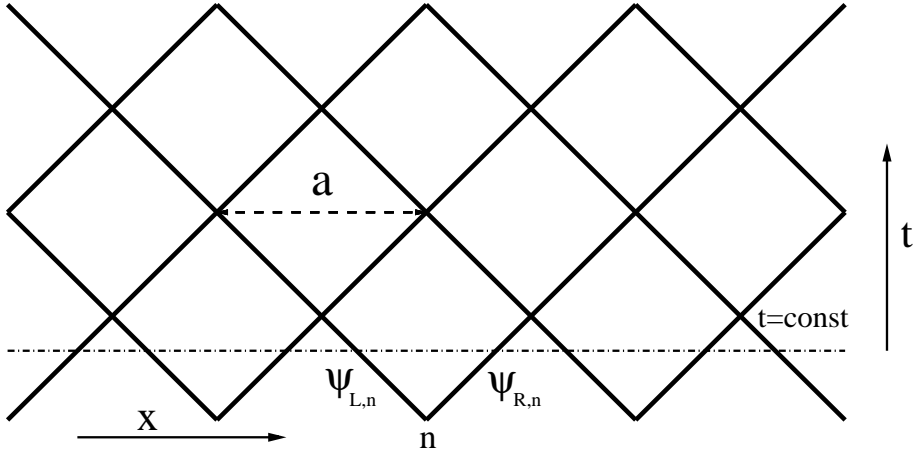}
\end{flushleft}
\caption{{\footnotesize
The pictorial representation of the light-cone lattice. 
}}
\label{figure}
\end{figure}

The light-cone lattice approach of \cite{ddvlc} provides an integrable lattice regularization of the MT model in the even topological charge sector of theory. 
In this description the space-time is discretized along the light-cone directions: $x_{\pm}=x\pm t$ with an even number of lattice sites in the spatial direction.
The sites of the light-cone lattice correspond to the discretized points of space-time. The left- and right-mover fermion fields live on the left- and right-oriented
edges of the lattice. In this manner a left- and a right-mover  fermion field can be assigned to each site of the lattice (See figure \ref{figure}.).

Lattice fermion fields satisfy the anticommutation relations:
 \begin{equation}
 \{\psi_{A,n},\psi_{B,m}\}=0, \quad \{\psi_{A,n},\psi^{+}_{B,m}\}=\delta_{AB}\, \delta_{nm}, 
\quad A,B=R,L, \quad 1\leq m,n\leq N.
 \end{equation}
 Then left- and right-mover fields live on the odd and even edges of of the light-cone lattice respectively:
 \begin{equation}
 \psi_{R,n}=\psi_{2 n}, \quad \psi_{L,n}=\psi_{2n-1}, \quad 1\leq n \leq \tfrac{N}{2}.
 \end{equation}
 In this regularization, the variables $\psi_n$ are used to formulate the model and they are related to the commonly used spin variables 
 by a Jordan-Wigner transformation:
 \begin{equation} \label{JWtf}
 \psi^+_n=\sigma^+_n \prod\limits_{l=1}^{n-1} \sigma^z_l, \qquad \psi_n=\sigma^-_n \prod\limits_{l=1}^{n-1} \sigma^z_l.
 \end{equation}
 The $U_L$ and $U_R$ light-cone evaluation operators of the model are given by inhomogeneous transfer 
matrices of the 6-vertex model with appropriate alternating inhomogeneities as follows.
 
 Let us consider the 6-vertex model with the following $R$-matrix:
 \begin{equation} \label{Rmatrix}
 R(\lambda)=\begin{pmatrix} 1 & 0 & 0 & 0 \\
 0 & \tfrac{\sinh(\lambda)}{\sinh(\lambda-i \gamma)} & \tfrac{\sinh(-i \gamma)}{\sinh(\lambda-i \gamma)} & 0 \\
0 & \tfrac{\sinh(-i \gamma)}{\sinh(\lambda-i \gamma)}  & \tfrac{\sinh(\lambda)}{\sinh(\lambda-i \gamma)} & 0 \\
0 & 0 & 0 & 1 
\end{pmatrix},
\end{equation}
 where $\lambda$ is the spectral parameter and $\gamma$ is the anisotropy parameter 
which encodes the coupling dependence of the MT model. 
 The coupling dependence of $\gamma$ is given by:
 \begin{equation}
\gamma=\frac{\pi}{p+1}, \qquad 0<p< \infty,
 \end{equation}
 where $p$ parameterizes the coupling constant of the SG and MT models by the formula{\footnote{This parameterization 
is introduced to relate our results easier to the DDV
 equation.}}:
 \begin{equation}
 \frac{\beta^2}{4 \pi}=\frac{1}{1+\tfrac{g}{4 \pi}}= \frac{2 p}{p+1}.
 \end{equation}
 The R-matrix (\ref{Rmatrix}) acts on the tensor product of two linear spaces both being isomorfic to $\mathbb{C}^2.$ 
 As usual, the  $R$-matrix acting on $V_1(\lambda_1) \otimes V_2(\lambda_2)$ is denoted by $R_{12}(\lambda_1-\lambda_2).$ 
The monodromy matrix acts on $V_0$ and the quantum space of the model 
 ${\cal{H}}={{\otimes}_{i=1}^N} \, V_i$ and is given by: 
 \begin{equation} \label{monodromy}
 T(\lambda|\vec{\xi})=R_{01}(\lambda-\xi_1)\, R_{02}(\lambda-\xi_2)\, ...R_{0N}(\lambda-\xi_N)=\begin{pmatrix} A(\lambda) & B(\lambda) \\
 C(\lambda) & D(\lambda) \end{pmatrix}_{[0]},
 \end{equation}
 where $\vec{\xi}$ is the $N$-dimensional inhomogeneity vector given by:
 \begin{equation}\label{2.11}
 \vec{\xi}=\{\xi_-,\xi_+,\xi_-,\xi_+,...,\xi_-,\xi_+\},
 \end{equation}
 with 
 \begin{equation} \label{inho}
 \xi_{\pm}=\pm \rho -i \tfrac{\gamma}{2}.
 \end{equation}
 Here the parameter $\rho$ is part of the regularization scheme.
This is why it depends on the lattice spacing or equivalently on the number of 
lattice sites. This dependence is given by the formula:
\begin{equation} \label{rho}
\rho=\tfrac{\gamma}{\pi} \, \ln \tfrac{4}{{\cal M} \, a}=\tfrac{\gamma}{\pi} \, \ln \tfrac{2 \, N}{{\cal M} \, L},
\end{equation}
where ${\cal M}$ is the physical mass of fermions (solitons), $a$ denotes the lattice spacing, $N$ is the 
number{\footnote{In this convention, in the light-cone lattice the number of lattice sites in spatial direction is $\tfrac{N}{2}.$ See figure \ref{figure}.}} 
of lattice sites of the 6-vertex model and $L$ is the volume. 
Due to the integrability of the model the transfer matrixes form a commutative family of operators on the quantum space of the model:
\begin{equation} \label{trans}
{\cal T}(\lambda|\vec{\xi})=\text{Tr}_0 \, T(\lambda|\vec{\xi}), \qquad \left[ {\cal T}(\lambda|\vec{\xi}),{\cal T}(\lambda'|\vec{\xi}) \right] =0.
\end{equation}
The $U_L$ and $U_R$ light-cone evaluation operators of the regularized MT model are given by the transfer matrices:
\begin{equation} \label{URL}
 U_L=e^{i\tfrac{2}{a}(H-P)}={\cal T}(\xi_+|\vec{\xi}), \qquad U_R^+=e^{-i\tfrac{2}{a}(H+P)}={\cal T}(\xi_-|\vec{\xi}),
\end{equation}
where $H$ is the Hamiltonian and $P$ is the momentum of the model. From this description it follows that the eigenstates of the Hamiltonian 
are the eigenvectors of the commuting transfer matrices. These eigenvectors can be obtained via the algebraic Bethe Ansatz technique \cite{FST79}.

\subsection{Algebraic Bethe Ansatz}

In the framework of algebraic Bethe Ansatz method, 
the eigenstates of the mutually commuting family of transfer matrices (\ref{trans})
 are constructed by acting with a product of $B$-operators 
on the reference state $|0 \rangle$, which is the completely ferromagnetic 
$S_z=\tfrac{N}{2}$ state of the model: 
\begin{equation} \label{sajatvec}
|\vec{\lambda}\rangle=|\lambda_1,\lambda_2,..,\lambda_m \rangle=B(\lambda_1)\, B(\lambda_2)\, ...B(\lambda_m)\,|0 \rangle, \qquad  S_z|\vec{\lambda}\rangle=(\tfrac{N}{2}-m)|\vec{\lambda}\rangle.
\end{equation}
Such a state is an eigenstate provided the spectral parameters in the argument of the $B$-operators satisfy the 
Bethe equations:
\begin{equation} \label{BAE}
\prod\limits_{i=1}^N \, \frac{\sinh(\lambda_a-\xi_i-i \gamma)}{\sinh(\lambda_a-\xi_i)} \, 
\prod\limits_{b=1}^m \, \frac{\sinh(\lambda_a-\lambda_b+i \gamma)}{\sinh(\lambda_a-\lambda_b-i \gamma)}=-1, \qquad a=1,...,m.
\end{equation}
The eigenvalues of the transfer matrices can also be expressed in terms of the Bethe-roots:
\begin{equation} \label{transsajat}
{\cal T}_{\vec{\lambda}}(\mu|\vec{\xi})=\prod\limits_{k=1}^m \, \frac{\sinh(\mu-\lambda_k+i \gamma)}{\sinh(\mu-\lambda_k)}+
\prod\limits_{i=1}^N \, \frac{\sinh(\mu-\xi_i)}{\sinh(\mu-\xi_i-i \gamma)} \,
\prod\limits_{k=1}^m \, \frac{\sinh(\mu-\lambda_k-i \gamma)}{\sinh(\mu-\lambda_k)}.
\end{equation}
The Bethe-equations can be reformulated in terms of the so-called counting-function $Z_\lambda(\lambda)$:
\begin{equation} \label{Zl}
(-1)^\delta \, e^{i \, Z_\lambda(\lambda_a)}=-1, \qquad \delta=m \, (\text{mod} \, 2),  \quad a=1,..,m,
\end{equation}
where 
\begin{equation} \label{uj}
(-1)^\delta \, e^{i \, Z_\lambda(\lambda)}=\prod\limits_{i=1}^N \, \frac{\sinh(\lambda-\xi_i-i \gamma)}{\sinh(\lambda-\xi_i)} \, 
\prod\limits_{b=1}^m \, \frac{\sinh(\lambda-\lambda_b+i \gamma)}{\sinh(\lambda-\lambda_b-i \gamma)}.
\end{equation}
For the proper definition of $Z_\lambda(\lambda)$ the logarithm of (\ref{uj}) should be taken, such that 
the counting function should be continuous along the real axis. 
This can be achieved by defining the function \cite{ddv97}:
\begin{equation} \label{phinu}
\phi_\nu(\lambda)=-i \, \log \frac{\sinh(i \tfrac{\gamma}{2} \nu -\lambda)}{\sinh(i \tfrac{\gamma}{2} \nu +\lambda)}, \qquad 0<\nu, \qquad \phi_{\nu}(0)=0, \qquad |\text{Im} \, \lambda| < \nu.
\end{equation}
 The function $\phi_\nu(\lambda)$ can be continued analytically to the 
 regime $|\text{Im} \, \lambda| > \nu$ by the requirements that its logarithmic discontinuities should 
run parallel to the real axis and it should be an odd function on the entire complex plane.
 Using this analytically continued $\phi_\nu(\lambda)$, the definition of the counting-function 
specified to the inhomogeneities (\ref{inho}) is given by the formula \cite{ddv97}:
\begin{equation} \label{countfv}
Z_\lambda(\lambda)=\frac{N}{2} \, \left( \phi_1(\lambda-\rho)+\phi_1(\lambda+\rho)\right)-\sum\limits_{k=1}^{m} \, \phi_2(\lambda-\lambda_k).
\end{equation}
Using $Z_\lambda(\lambda)$, the Bethe-equations (\ref{BAE}) can be reformulated in their logarithmic form by the formula:
\begin{equation} \label{BAEZ}
Z_{\lambda}(\lambda_a)=2 \pi \, I_a, \quad  \quad I_a\in \mathbb{Z}+\tfrac{1+\delta}{2} \qquad a=1,..,m.
\end{equation}
We note that the role of $\delta$ is to determine whether the quantum numbers $I_a$ should be integers or half-integers. 
 The vacuum of the field theory corresponds to the $\delta=0, \, S_z=0,$ antiferromagnetic vacuum of the 
lattice-model{\footnote{According to (\ref{Zl}), the $\delta=0$ requirement implies that  
$\tfrac{N}{2}$ must be even on the lattice.}}. 
This is formed by $N/2$ real Bethe-roots, such that to all quantum numbers satisfying the inequality
$Z_{\lambda}(-\infty)\leq 2 \pi \, I_a  \leq Z_{\lambda}(\infty)$ there exist a real Bethe-root in (\ref{BAEZ}).
The excitations above this vacuum are characterized by complex Bethe-roots and holes, where holes are such real solutions of
 (\ref{Zl}), which are not Bethe-roots. In the logarithmic form of the equations quantum numbers can be 
assigned to holes as well:
\begin{equation} \label{BAEH}
Z_{\lambda}({h}_k)=2 \pi \, I_k, \quad  \quad I_k\in \mathbb{Z}+\tfrac{1+\delta}{2} \qquad k=1,..,m_H,
\end{equation} 
where $h_k$ denotes the positions of the holes and their number is denoted by $m_H.$

\subsection{The DDV equations}

The DDV equations{\footnote{A detailed review on the DDV equations can be found in \cite{Fevphd}.}} \cite{KP1}-\cite{FRT3} reformulate the 
Bethe-equations (\ref{BAE}) in terms of a set of nonlinear-integral equations,
 such that only those objects enter the equations, which characterize the excitations. In this paper we will compute 
diagonal form factors in the pure soliton sector of the theory, thus we recall here the form of the DDV equation 
only for the pure soliton- or equivalently for pure hole states. Here we present the equations in rapidity variables
 i.e. $\theta=\tfrac{\pi}{\gamma}\lambda$, 
because of two reasons. First, this way it is easier to find connection to the literature of the DDV equation \cite{ddv92}-\cite{FRT3},
 and on the other hand at the stage of our final results it is better to work in this convention, since in the field 
 theory this variable corresponds to the rapidity of particles. We recall the DDV equation for both the 
lattice and for
the continuum theories.
To do so, first we relate the lattice counting function in rapidity variables to $Z_{\lambda}(\lambda)$ of  (\ref{countfv}). The relation is given by $Z_N(\theta)=Z_{\lambda}(\tfrac{\gamma}{\pi}\theta).$ 
The DDV equation for $Z_N(\theta)$ in the pure hole sector reads as:
\begin{equation} \label{DDVlat}
\begin{split}
Z_N(\theta)=\frac{N}{2}\left\{\arctan\left[\sinh(\theta-\Theta)\right]+\arctan\left[ \sinh(\theta+\Theta)\right]  \right\}+\sum\limits_{k=1}^{m_H} \, \chi(\theta-H_k) \\
+ \, \int\limits_{-\infty}^{\infty} \frac{d\theta'}{2 \pi i} \, G(\theta-\theta'-i\eta) \, L_N^{(+)}(\theta'+i \eta)
-\, \int\limits_{-\infty}^{\infty} \frac{d\theta'}{2 \pi i} \, G(\theta-\theta'+i\eta) \, L_N^{(-)}(\theta'-i \eta),
\end{split}
\end{equation}
where $\chi(\theta)$ is the soliton-soliton scattering 
phase and $G(\theta)$ is its derivative:
\begin{equation} \label{G}
G(\theta)=-i \, \frac{d}{d\theta} \log S_{++}^{++}(\theta)=\! \int\limits_{-\infty}^{\infty} \! d\omega \, 
e^{-i \, \omega \theta} \, \frac{\sinh(\tfrac{(p-1) \,\pi \omega}{2})}{2 \cosh( \tfrac{\pi \omega}{2}) 
\,\sinh(\tfrac{p \,  \pi \, \omega}{2})},
\end{equation}
$0<\eta<\text{min}(p \pi,\pi)$ is an arbitrary positive contour-integral parameter, which must be smaller than 
the distance of the first pole of $G(\theta)$ from the real axis. Furthermore, $L_N^{(\pm)}(\theta)$ denotes the 
nonlinear combinations of $Z_N(\theta)$:
\begin{equation} \label{LNpm}
L_N^{(\pm)}(\theta)=\ln\left(1+(-1)^\delta \, e^{\pm i \, Z_N(\theta)} \right),
\end{equation}
$\Theta=\ln \tfrac{2 \, N}{{\cal M} \, L}$ is the inhomogeneity parameter and $H_k=\tfrac{\pi}{\gamma} h_k$ denote the 
positions of the holes in the rapidity convention. They are subjected to the quantization equations:
\begin{equation} \label{QHk}
Z_N(H_k)=2 \pi \, I_k\, \qquad I_k\in \mathbb{Z}+\tfrac{1+\delta}{2} \qquad k=1,..,m_H.
\end{equation}
A counting-equation \cite{ddv97} can be derived, which tells us how the number of excitation characterizing objects 
is related to the spin or equivalently to the conserved quantum number of the state.
For pure hole states without special objects{\footnote{Special objects are points on the complex plane, where 
the $L_N^{(\pm)}(\theta)$ jumps along the integration contour due to going though the branch cut of the logarithm.
For more detail see for example \cite{ddv97,Fevphd}.} the counting-equation on the lattice takes the form:
\begin{equation} \label{countlat}
m_H=2 \, S_z-2 \left[ \tfrac{1}{2}+\tfrac{S_z}{p+1} \right],
\end{equation}
where here $[...]$ stands for integer part. Since $S_z=\tfrac{N}{2}-m$, this equation tells us that, 
on a lattice with even number of sites, only states with even number of holes exist. 
The lattice counting-function $Z_N(\theta)$ depends on the number of lattice sites $N.$ 
It has a continuum limit, which is just its $N \to \infty$ limit \cite{ddv92,ddv95}:
\begin{equation} \label{Zcontlim}
Z(\theta)=\lim_{N \to \infty} Z_N(\theta), \qquad
L_{\pm}(\theta)=\lim_{N \to \infty}\, L_N^{(\pm)}(\theta)=\ln\left(1+(-1)^\delta \, e^{\pm i \, Z(\theta)} \right).
\end{equation}
With these notations the continuum DDV equations are just the $N \to \infty$ limit of the lattice ones (\ref{DDVlat}):. 
\begin{equation} \label{DDVcont}
\begin{split}
Z(\theta)=\ell \sinh \theta +\sum\limits_{k=1}^{m_H} \, \chi(\theta-H_k) 
+ \, \int\limits_{-\infty}^{\infty} \frac{d\theta'}{2 \pi i} \, G(\theta-\theta'-i\eta) \, L_{+}(\theta'+i \eta) \\
-\, \int\limits_{-\infty}^{\infty} \frac{d\theta'}{2 \pi i} \, G(\theta-\theta'+i\eta) \, L_-(\theta'-i \eta),
\end{split}
\end{equation}
where $\ell={\cal M} \, L$ with $L$ being the volume and ${\cal M}$ is the soliton mass.
The energy and momentum of these hole states in the continuum read as:
\begin{equation} \label{Econt}
\begin{split}
E={\cal M} \, \sum_{k=1}^{m_H} \, \cosh H_k -\frac{{\cal M}}{2 \pi i} \sum\limits_{\alpha=\pm} \,
\int\limits_{-\infty}^{\infty} d\theta \, \sinh(\theta+i \, \alpha\, \eta) \, L_{\alpha}(\theta+i \, \alpha \, \eta),
\end{split}
\end{equation}
\begin{equation} \label{Pcont}
\begin{split}
P={\cal M} \, \sum_{k=1}^{m_H} \, \sinh H_k -\frac{{\cal M}}{2 \pi i} \sum\limits_{\alpha=\pm} \,
\int\limits_{-\infty}^{\infty} d\theta \, \cosh(\theta+i \, \alpha\, \eta) \, L_{\alpha}(\theta+i \, \alpha \, \eta).
\end{split}
\end{equation}
Since in the large volume limit $L_{\alpha}(\theta+i \, \alpha \, \eta) \to 0$, 
from (\ref{Econt}) and (\ref{Pcont}) it can be seen that in the large volume limit the holes correspond to the 
rapidities of the solitons. This is why in the sequel we will refer to holes as solitons.
It also turns out \cite{ddv97} that the counting equation (\ref{countlat}) changes in the continuum and it reads{\footnote{For pure soliton states without special objects.}}:
\begin{equation} \label{contcont}
Q=m_H,
\end{equation} 
where $Q$ is the $U(1)$ (topological) charge of the continuum model.

The choice{\footnote{On the lattice the actual value of $\delta$ can be influenced by the parity of $\tfrac{N}{2}.$}} of $\delta$ is 
crutial in the continuum theory. In the even charge sector of the theory 
$\delta=0.$ In the odd charge sector the choice $\delta=0$ corresponds to the MT fermions, while the $\delta=1$ 
choice describes the SG solitons \cite{FRT1}-\cite{FRT3}.

Though in this paper we will make computations only in the twistless case, which can describe only the even 
topological charge sector of the model, we note that in \cite{BHsg} it has been shown that the odd charge sector 
can also be investigated from the lattice, if the 6-vertex model with an twist angle $\omega=\tfrac{\pi}{2}$
 is considered.

\subsection{The $U(1)$ current in spin variables}

Our purpose is to compute the finite volume form-factors of local operators of the MT/SG models in the
framework of QISM.  To achieve this plan, the first step is to relate the lattice operators to the continuum ones.
Due to renormalization effects, this is a complicated task in general. To avoid complications{\footnote{Here we mostly 
think of operator mixing.}} coming from 
renormalization effects we will restrict our attention to operators related to the $U(1)$ symmetry of the 
model. The $U(1)$ symmetry is present in both the lattice and the continuum theories, thus it is plausible to assume 
that the $U(1)$ conserved charge of the lattice theory is mapped to the $U(1)$ charge of the continuum theory.

The counting equations (\ref{countlat}) and (\ref{contcont}) suggest{\footnote{At least for large enough 
values of $p$, when the integer part becomes zero.}} the $Q \sim 2 S_z$ identification
 between the lattice and continuum conserved quantities. 
 This helps us to define the correct normal ordering for the lattice fermion fields as follows.
Assuming the $Q \sim 2 S_z$  relation, the lattice topological charge can be expressed by lattice fermion 
fields using a Jordan-Wigner transformation 
(\ref{JWtf}):
\begin{equation} \label{QSz}
\begin{split}
Q\sim2\, S_z=\sum\limits_{n=1}^N \, \sigma^z_n=\sum\limits_{n=1}^N \, \left( \psi^+_n \psi_n-\tfrac{1}{2} \right).
\end{split}
\end{equation}
In the continuum theory the topological charge is given by the integral:
\begin{equation} \label{Qint00}
\begin{split}
Q=\int\limits_{0}^L dx \, \left( :\Psi_R^+ \Psi_R:(x)+ :\Psi_L^+ \Psi_L:(x)\right),
\end{split}
\end{equation}
which can be approximated on the lattice by the discrete sum:
\begin{equation} \label{Qlat0}
\begin{split}
Q\approx \sum\limits_{n=1}^{\tfrac{N}{2}} \, \left( :\psi_{2n}^+ \psi_{2n}:+ :\psi_{2n-1}^+ \psi_{2n-1}:\right).
\end{split}
\end{equation}
The comparison of (\ref{QSz}) and (\ref{Qlat0}) offers a natural definition for the normal ordering of lattice 
Fermi-fields:
\begin{equation} \label{fno}
:\psi^+_n \psi_n:=\psi^+_n \psi_n-\tfrac{1}{2}=\sigma^z_n.
\end{equation}  
Our purpose is to determine the lattice counterparts of the conserved current belonging to the $U(1)$ current of 
the MT model. The index-0 component of the current is the charge density. This can be computed from (\ref{QSz}) and
(\ref{fno}):
\begin{equation} \label{j0latt}
\begin{split}
Q\sim2\, S_z=\sum\limits_{n=1}^{\tfrac{N}{2}} \, (\sigma^z_{2n}+\sigma^z_{2n-1})=
\sum\limits_{n=1}^{\tfrac{N}{2}} \,a \, \frac{\sigma^z_{2n}+\sigma^z_{2n-1}}{a}=
\sum\limits_{n=1}^{\tfrac{N}{2}} \,a \cdot j_0(n a)\to \int\limits_{0}^L \! dx \, j_0(x),
\end{split}
\end{equation} 
where $a=\tfrac{2 \, L}{N}$ is the lattice constant and the index-$0$ component of the current at the lattice sites 
can be expressed in terms of lattice spin variables as:
\begin{equation} \label{j0na}
j_0(n a)=\frac{\sigma^z_{2n}+\sigma^z_{2n-1}}{a}=\frac{N}{L}\frac{\sigma^z_{2n}+\sigma^z_{2n-1}}{2}.
\end{equation}
In the continuum theory the conserved current is given by:
\begin{equation} \label{Jmu}
J_\mu=:\bar{\Psi}\, \gamma_\mu \Psi:, \qquad \mu=0,1.
\end{equation}
which can be written in component fields as:
\begin{equation} \label{J01}
\begin{split}
J_0 &=:\Psi_R^+ \Psi_R:+:\Psi_L^+ \Psi_L:, \\
J_1 &=:\Psi_R^+ \Psi_R:-:\Psi_L^+ \Psi_L:.
\end{split}
\end{equation}
Since left- and right-mover fields live on the odd and even links of our lattice respectively, comparing 
(\ref{j0na}) and (\ref{J01}) gives immediately the index-$1$ component of the current in terms of spin variables:
\begin{equation} \label{j1na}
j_1(n a)=\frac{\sigma^z_{2n}-\sigma^z_{2n-1}}{a}=\frac{N}{L}\frac{\sigma^z_{2n}-\sigma^z_{2n-1}}{2}.
\end{equation}
Consequently (\ref{j0na}) and (\ref{j1na}) indicates that the computation of form factors of the current 
$J_\mu$ is reduced to compute form factors of $\sigma^z_n$ on the lattice. This  can be achieved 
within the framework of QISM \cite{KMT99,KMT00}.

We close this section with an important remark concerning the continuum limit and our notations. 
It can be recognized, that as far as the notation is concerned, we made difference between the continuum and the 
lattice notations of the $U(1)$ current. Namely,  $J_\mu(x)$ denotes the current in the continuum field 
theory, while $j_\mu(x)$ denotes the lattice analog of the continuum current, the derivation of which was based on 
the identification of the topological charge of the continuum field theory with $2 S_z$ of the corresponding lattice 
 theory. However, in section \ref{solutions}. it will turn out that 
the two quantities are not equal, but only proportional. 
We just anticipate their relation, which is given by the formula (\ref{Jmuop}):
\begin{equation} \label{Jjmu}
\begin{split}
J_\mu(x)=\tfrac{p}{p+1} \, j_\mu(x), \qquad \mu=0,1.
\end{split}
\end{equation}
We note, that the renormalization factor $\tfrac{p}{p+1}$ tends to $1$, as $p$ tends to infinity in accordance with 
the indication of (\ref{countlat}).

\section{Form-factors in the QISM framework}

In the previous section we argued that the computation of form-factors of the $U(1)$ current of the MT/SG
model is equivalent to the determination of the form-factors of $\sigma^z_n$ on the lattice.
Our approach to compute the finite volume form-factors of local operators having lattice counterparts, consists of two steps;
first one should compute the form-factors on the lattice. The result will depend on the number 
of lattice sites $N$. 
Then the $N \to \infty$ limit of the lattice result gives the required result for the continuum theory{\footnote{In many cases
the careful analysis of renormalization constants is also necessary.}}.
 We will demonstarate, that this method works fine by the computation of the diagonal matrix elements 
of the $U(1)$ current. The details of the computations enlight, that the diagonal matrix elements of 
other combinations of local Fermi fields and their derivatives can also be computed by this method. 

Moreover, this procedure gives a theoretical framework also for the computation of 
nondiagonal matrix elements of the operators. 

In this section we collect the most important formulas being necessary for the computations. 
Consider the following vector of the Hilbert-space:
\begin{equation} \label{bra}
|\vec{\lambda} \rangle=B(\lambda_1)\, B(\lambda_2)...B(\lambda_m) \, |0 \rangle.
\end{equation}
This is called Bethe-state if the numbers $\lambda_j$ are arbitrary and is called Bethe-eigenstate
if all $\lambda_j$ are solutions of the Bethe equations (\ref{BAE}).
Then the corresponding "bra" vector is given by:
\begin{equation} \label{ket}
\langle \vec{\lambda}|=\langle 0|\, C(\lambda_m)... C(\lambda_2) \,C(\lambda_1).
\end{equation}
To get all form-factors of the current, one should be able to compute the lattice form factors:
\begin{equation} \label{ff0}
\frac{\langle \vec{\mu}|\sigma^z_n|\vec{\lambda}\rangle}
{\sqrt{\langle \vec{\lambda} |\vec{\lambda}\rangle \, \langle \vec{\mu} |\vec{\mu} \rangle}},
\end{equation}
where both $|\vec{\mu} \rangle$ and $|\vec{\lambda}\rangle$ are Bethe-eigenstates, but in this paper
we will focus on computing only the diagonal matrix elements:
\begin{equation} \label{ffd}
\langle \sigma^z_n \rangle_\lambda=\frac{\langle \vec{\lambda}|\sigma^z_n|\vec{\lambda}\rangle}
{\langle \vec{\lambda} |\vec{\lambda}\rangle },
\end{equation}
which turns out to be a much simpler problem.

Here we recall the most important formulas \cite{KMT00}, which are necessary to perform our calculations.
For the computations  only the Yang-Baxter algebra and the elements of the monodromy matrix (\ref{monodromy}) 
are used, this is why it is important to express the local spin operators in terms of the $A,B,C,D$ operators
 of the monodromy matrix (\ref{monodromy}). It has been done in \cite{KMT99} and the relations are summarized by the
formula:
\begin{equation} \label{ISM}
E^{ab}_n=\prod\limits_{i=1}^{n-1} \, (A+D)(\xi_i) \, \, T_{ab}(\xi_n) \, \prod\limits_{i=n+1}^{N} \, 
(A+D)(\xi_i), \qquad a,b=1,2,
\end{equation}
where the operator $E_n$ is given in terms of spin operators as follows:
\begin{equation} \label{En}
E^{11}_n=\frac{1}{2}({1}_n+\sigma^z_n), \quad E_n^{12}=\sigma^-_n, \quad E_n^{21}=\sigma^+_n, \quad E^{22}_n=\frac{1}{2}({1}_n-\sigma^z_n).
\end{equation}
In our actual computations we use the $22$-component of (\ref{ISM}):
\begin{equation} \label{e22}
e_n=\frac{1}{2}({1}_n-\sigma^z_n)=\prod\limits_{i=1}^{n-1} \, (A+D)(\xi_i) \, \, D(\xi_n) \, \prod\limits_{i=n+1}^{N} \, 
(A+D)(\xi_i),
\end{equation}
where for short we introduced the notation $e_n=\frac{1}{2}({1}_n-\sigma^z_n).$
We compute the expectation values of $e_n$ on the lattice, since apart from a trivial constant and sign it is
equal to the required matrix element $\tfrac{1}{2}\langle \sigma^z_n \rangle_\lambda$. 
We note, that the lattice part of our computations is a special case of the computations done in \cite{KMT00} for the emptiness formation probability. 
This is why we will mostly use the logic and formulas of \cite{KMT00}.

To compute $\langle e_n \rangle_\lambda$ from (\ref{e22}), one should know how the operator $D(\xi_n)$ 
acts{\footnote{We just note that the factors coming from the $(A+D)$-wings of (\ref{e22}) give 
scalar factors since the sandwiching states are eigenstates of $(A+D)$.}} on the
 "bra"-vector (\ref{ket}). This is given by the following formula \cite{KMT00}:
\begin{equation} \label{Dact}
\langle 0|\prod\limits_{k=1}^m \, C(\lambda_k) \, D(\xi_n)=\! \! \sum\limits_{a=1}^m \! \frac{1}{r(\lambda_a)} \,
\frac{\prod\limits_{k=1}^m \sinh(\lambda_a-\lambda_k-i \, \gamma)}
{ \sinh(\lambda_a-\xi_n)\! \prod\limits_{k=1 \atop k \neq a 
}^m \sinh(\lambda_a-\lambda_k) } \,
\langle 0|\! \prod\limits_{ k=1 \atop k \neq a
 }^m  C(\lambda_k) \, C(\xi_n),
\end{equation}
where we explicitely exploited that $\xi_n$ is one of the inhomogeneities of the vertex model and
introduced:
\begin{equation} \label{rdef}
r(\lambda)=\prod\limits_{j=1}^N \, \frac{\sinh(\lambda-\xi_j-i \,\gamma)}{\sinh(\lambda-\xi_j)}.
\end{equation}
As a consequence of (\ref{rdef}) and (\ref{BAE}) it satisfies the identities:
\begin{equation} \label{rid}
\prod\limits_{k=1}^m r(\lambda_k)=1, \qquad \frac{1}{r(\xi_j)}=0, \quad j=1,...,N.
\end{equation}
The last ingredient necessary for the computations is the scalar product a Bethe-state and a Bethe-eingenstate.
Let $|\mu \rangle$ an arbitrary Bethe-state in the sense of (\ref{bra}) and $|\lambda \rangle$ be a Bethe-eigenstate.
Then their scalar product is given by the formula \cite{Sla89}:
\begin{equation} \label{skalarH}
\langle \vec{\mu}|\vec{\lambda} \rangle=\langle \vec{\lambda} |\vec{\mu} \rangle=
\prod\limits_{l=1}^N \, \frac{1}{ r(\mu_l)} \cdot
\frac{\text{det} \, H(\vec{\mu}|\vec{\lambda})}{ \prod\limits_{j>k} \sinh(\mu_k-\mu_j) \, \sinh(\lambda_j-\lambda_k)},
\end{equation}
where $H(\vec{\mu}|\vec{\lambda})$ is an $m \times m$ matrix with entries:
\begin{equation} \label{Hdef}
H_{ab}(\vec{\mu}|\vec{\lambda})=\frac{\sinh(-i \, \gamma)}{\sinh(\lambda_a-\mu_b)} 
\left( r(\mu_b) \, \frac{\prod\limits_{k=1}^m  \sinh(\lambda_k-\mu_b-i \, \gamma)}{\sinh(\lambda_a-\mu_b-i \, \gamma)} 
-\frac{\prod\limits_{k=1}^m  \sinh(\lambda_k-\mu_b+i \, \gamma)}{\sinh(\lambda_a-\mu_b+i \, \gamma)}
\right).
\end{equation}
The special case of the formula (\ref{skalarH}), when both states correspond to the same Bethe-eigenvector{\footnote{If the two eingenstates are different the scalar product is zero.}},
 gives the Gaudin formula:
\begin{equation} \label{Gaudin}
\langle \vec{\lambda} |\vec{\lambda} \rangle  =
\frac{\prod\limits_{j=1}^{m} \prod\limits_{k=1}^{m} \sinh(\lambda_j-\lambda_k-i \, \gamma)
 }{ \prod\limits_{j>k} \sinh(\lambda_k-\lambda_j) \, \sinh(\lambda_j-\lambda_k)} \cdot \text{det} \, \Phi(\vec{\lambda}),
\end{equation}
where  
 $\Phi(\vec{\lambda})$ is the Gaudin-matrix, which can be obtained from the counting-function 
(\ref{countfv}) as follows:
\begin{equation} \label{Phi}
\Phi_{ab}(\vec{\lambda})=-i \,\frac{\partial}{\partial \lambda_b} \, Z_{\lambda}(\lambda_a|\vec{\lambda}), \quad a,b=1,..,m,
\end{equation}
where we indicated, that the counting-function should be considered as a function of the Bethe-roots.
 This Bethe-root dependence can be read off from (\ref{countfv}).

From (\ref{Hdef}) it can be seen, that the matrix element $H_{ab}(\vec{\mu}|\vec{\lambda})$ depend on only 
one single component of the vector $\vec{\mu}.$ This observation makes it possible remarkable  simplifications, 
when diagonal form-factors are computed. In this case 
 one needs to compute scalar 
products, when the components of the vector $\vec{\mu}$ take values either from the set of Bethe-roots 
$\{\lambda_j\}_{j=1,..m}$ or from the set of inhomogeneities $\{\xi_k\}_{k=1,..N}$ of the model. 
In these cases the matrix elements of $H(\vec{\mu}|\vec{\lambda})$ take the form:
\begin{equation} \label{HPhi}
H_{ab}(\vec{\mu}|\vec{\lambda})\big|_{\mu_b \to \lambda_c}=(-1)^{m-1} \, \prod\limits_{j=1}^m \, 
\sinh(\lambda_c-\lambda_j-i \, \gamma) \, \Phi_{ac}(\vec{\lambda}), \quad a,b,c=1,..,m.
\end{equation}
 \begin{equation} \label{Hxi}
\frac{1}{r(\mu_b)}\, H_{ab}(\vec{\mu}|\vec{\lambda})\big|_{\mu_b \to \xi_c}=
\frac{(-1)^{m} \,\sinh(-i\, \gamma) \prod\limits_{j=1}^m \, \sinh(\xi_c-\lambda_j+i \, \gamma)}
{\sinh(\lambda_a-\xi_c) \, \sinh(\lambda_a-\xi_c-i \, \gamma)}, \quad a,b,c=1,..,m.
\end{equation}

\section{The computation of $\langle e_n \rangle_\lambda$ }

Now we are in the position to compute $\langle e_n \rangle_\lambda$ on the lattice. First the contribution of 
the eigenvalues of the transfer-matrices are lifted:
\begin{equation} \label{e_1}
\langle e_n \rangle_\lambda=\prod\limits_{k=1}^{m}\, \frac{\sinh(\xi_n-\lambda_k)}{\sinh(\xi_n-\lambda_k+i \, \gamma)} 
\cdot \langle D(\xi_n) \rangle_\lambda.
\end{equation}
As a consequence of (\ref{Dact}) the expectation value $\langle D(\xi_n) \rangle_\lambda$ can be written as: 
\begin{equation} \label{Dexp}
\langle D(\xi_n) \rangle_\lambda=\frac{\langle \vec{\lambda}|D(\xi_n)|\vec{\lambda}\rangle}{\langle \vec{\lambda}|\vec{\lambda} \rangle}=
 \sum\limits_{A=1}^m \! \frac{1}{r(\lambda_A)} \,
\frac{\prod\limits_{k=1}^m \sinh(\lambda_A-\lambda_k-i \, \gamma)}
{ \sinh(\lambda_A-\xi_n)\! \prod\limits_{ k=1 \atop k\neq A
}^m \sinh(\lambda_A-\lambda_k) } \cdot \frac{\langle \vec{\mu}^{(A)}|\vec{\lambda} \rangle}{\langle \vec{\lambda}|\vec{\lambda} \rangle},
\end{equation}
where $\vec{\mu^{(A)}}$ is an $m$-component vector,
 which differs from $\vec{\lambda}$ only in its $A$th component, which is
equal to the inhomogeneity corresponding to the $n$th site:
\begin{equation} \label{muA}
\mu^{(A)}_k=\left\{
\begin{array}{c}  \lambda_k, \quad k \neq A, \\
\xi_n, \quad k=A.
\end{array} \right. \qquad \qquad k=1,..,m.
\end{equation}
Due to (\ref{HPhi}) and (\ref{Hxi}) using some simple determinant identities, 
$\langle \vec{\mu}^{(A)}|\vec{\lambda} \rangle$ can be written as:
 \begin{equation} \label{W_A}
\langle \vec{\mu}^{(A)}|\vec{\lambda} \rangle=\prod\limits_{j=1}^{m} \frac{\sinh(\xi_n-\lambda_j+i \, \gamma)}{\sinh(\xi_n-\lambda_j-i \, \gamma)} \cdot
\frac{\prod\limits_{b=1}^{m} \prod\limits_{j=1}^{m} \sinh(\mu_b^{(A)}-\lambda_j-i \, \gamma) \cdot \text{det} \, \hat{\cal H}(\vec{\mu}^{(A)}|\vec{\lambda})}
{ \prod\limits_{j>k}^{m} \sinh(\mu_k^{(A)}-\mu_j^{(A)}) \, \sinh(\lambda_j-\lambda_k) },
\end{equation}
where the $m \times m$ matrix $\hat{\cal H}(\vec{\mu}^{(A)}|\vec{\lambda})$ is given by:
\begin{equation} \label{Hhat}
\hat{\cal H}_{ab}(\vec{\mu}^{(A)}|\vec{\lambda})=\left\{
\begin{array}{l}
\tilde{\Phi}_{ab}(\vec{\lambda}), \qquad \qquad \qquad \qquad \qquad \quad \, \,  b \neq A, \\
{\cal V}_a \equiv
\frac{-\sinh(-i \, \gamma)}{\sinh(\lambda_a-\xi_n) \, \sinh(\lambda_a-\xi_n-i \,\gamma)}, \qquad b=A.
\end{array}
\right.
\end{equation}
Here 
\begin{equation} \label{Phikigyo}
\tilde{\Phi}_{ab}(\vec{\lambda})={\Phi}_{ab}(\vec{\lambda})\, \frac{1}{r(\lambda_b)}, \quad a,b=1,..,m.
\end{equation}
As a consequence of (\ref{rid}) $\text{det} \, {\Phi}(\vec{\lambda})=\text{det} \, \tilde{\Phi}(\vec{\lambda})$, thus 
in (\ref{Gaudin}) the ${\Phi}(\vec{\lambda}) \to \tilde{\Phi}(\vec{\lambda})$ replacement can be done. 
Using (\ref{muA}), (\ref{W_A}) and (\ref{Hhat}) one obtains:
\begin{equation} \label{Wtilde_A}
\frac{\langle \vec{\mu}^{(A)}|\vec{\lambda} \rangle}{\langle \vec{\lambda}|\vec{\lambda} \rangle}= \! 
 \prod\limits_{ j=1 \atop j \neq A
}^{m}  \frac{\sinh(\lambda_A-\lambda_j)}{\sinh(\xi_n-\lambda_j)} \cdot
\prod\limits_{j=1}^{m} \frac{\sinh(\xi_n-\lambda_j+i \, \gamma)}{\sinh(\lambda_A-\lambda_j-i \, \gamma)} \,
r(\lambda_A) \, \left( \Phi^{-1}(\vec{\lambda}) \cdot \hat{\cal H}(\vec{\mu}^{(A)}|\vec{\lambda}) \right)_{AA},
\end{equation}
where apart from simplifying the multiplicative factors, we used (\ref{Gaudin}) with the $\Phi \to \tilde{\Phi}$ 
replacement and computed the ratio of the determinants of $\hat{\cal H}(\vec{\mu}^{(A)}|\vec{\lambda})$ and $\tilde{\Phi}(\vec{\lambda})$ as the determinant of $\tilde{\Phi}^{-1}(\vec{\lambda}) \cdot \hat{\cal H}(\vec{\mu}^{(A)}|\vec{\lambda})$. As a consequence of (\ref{Hhat}), the latter matrix differs from the unity
 matrix only in its $A$th column. Thus its determinant could also be computed as an expression of matrix elements
of $\tilde{\Phi}^{-1}(\vec{\lambda})$ and $\hat{\cal H}(\vec{\mu}^{(A)}|\vec{\lambda})$.

Inserting (\ref{Wtilde_A}) into (\ref{Dexp}) and the result into (\ref{e_1}) one ends up with the simple result:
\begin{equation} \label{e_2}
\langle e_n \rangle_\lambda=-\sum\limits_{A=1}^m \left( \Phi^{-1}(\vec{\lambda}) \cdot \hat{\cal H}(\vec{\mu}^{(A)}|\vec{\lambda}) \right)_{AA}.
\end{equation}
Using (\ref{Hhat}), this can be written in components as:
\begin{equation} \label{e_3}
\langle e_n \rangle_\lambda=-\sum\limits_{a=1}^m \sum\limits_{b=1}^m  \Phi^{-1}_{ab}(\vec{\lambda}) 
\, {\cal V}_b=-\sum\limits_{a=1}^m S_a,
\end{equation}
where $S_a$ is the solution of the set of linear equations:
\begin{equation} \label{Sa}
\sum\limits_{b=1}^m \Phi_{ab}(\vec{\lambda}) \, S_b={\cal V}_a, \qquad a=1,..,m.
\end{equation}

\subsection{The determination of  $S_a$}\label{detSa}

In this subsection we show that equation (\ref{Sa}) for the vector $S_a$ can be formulated as a  
set of linear-integral-equations containing the counting-function of the model. The advantage of this 
formulation is that it allows one to take the continuum limit in a straightforward manner.

The first step is to compute the matrix elements of $\Phi(\vec{\lambda})$ from (\ref{Phi}):
\begin{equation} \label{Phiab}
\Phi_{ab}(\vec{\lambda})=-i \, Z^{\prime}_{\lambda}(\lambda_a) \, \delta_{ab}-2 \pi \, i \, K(\lambda_a-\lambda_b|\gamma), \quad a,b=1,..m,
\end{equation}
where
\begin{equation} \label{K1}
K(\lambda|\gamma)=\frac{1}{2 \pi} \frac{\sin(2 \, \gamma)}{\sinh(\lambda-i \, \gamma)\, \sinh(\lambda+i \, \gamma)}.
\end{equation}
Now an important remark is in order. From (\ref{Phiab}) it can be seen, that apart form the $\delta_{ab}$ term,
 the $ab$ matrix element of $\Phi(\vec{\lambda})$ is given by a function of two 
variables taken at the arguments $\lambda_a$ and $\lambda_b$, and similarly 
from (\ref{Hhat}) it is obvious that ${\cal V}_a$ is
an analytic function taken at the position $\lambda_a$. This suggests that the components of the unknown vector $S_a$ 
should be sought in the following form:
\begin{equation} \label{SXa}
S_a=X(\lambda_a), \quad a=1,...,m,
\end{equation}
where $X(\lambda)$ is supposed to be an analytic (meromorphic) function on the complex-plane{\footnote{The thermodynamic limit for the ground state
expectation value was treated by the same tacit assumption in \cite{KMT00}.}}. 
The advantage of such an Ansatz becomes obvious in the large $N$-limit, because summation for the 
large number of components becomes a convolution integral plus a remnant sum{\footnote{The convolution integral
 comes from the "Dirac-see" of real roots and the remnant sum is related to finite number of excitations above 
this see.}}.

To transform the sum in (\ref{Sa}) into an integral we should use the following lemma \cite{ddv95,ddv97}.
\newline {\bf Lemma:}
{ \it Let $\{\lambda_j\}_{j=1,..,m}$ solutions of the Bethe-equations (\ref{BAE})
 and let $f(\lambda)$ a meromorphic function, which is integrable on the real axis. 
Denote $p^{(f)}$ its pole located the closest to the real axis.
Then for $|\text{Im}\,\mu|<|\text{Im} \, p^{(f)}|$ the following equation
 holds:
\begin{equation} \label{sumrule}
\begin{split}
\sum\limits_{j=1}^{m} f(\mu-\lambda_j)&=\sum\limits_{j=1}^{m_C}f(\mu-c_j)-\sum\limits_{j=1}^{m_H} f(\mu-h_j)+
\int\limits_{-\infty}^{\infty} \!  \frac{d \lambda}{2 \pi} \, f(\mu-\lambda) \, Z'_\lambda(\lambda)- \\
&-\sum\limits_{\alpha=\pm} \int\limits_{-\infty}^{\infty} \! \frac{d \lambda}{2 \pi}  f(\mu-\lambda+i \, \alpha \, \eta) \,Z'_\lambda(\lambda+i \, \alpha \, \eta) \, 
{\cal F}^{(\lambda)}_{\alpha}(\lambda+i \, \alpha \, \eta),
\end{split}
\end{equation}
where 
\begin{equation} \label{calFl}
{\cal F}^{(\lambda)}_{\pm}(\lambda)=\frac{(-1)^\delta \, e^{\pm i \, Z_\lambda(\lambda)}}
{1+(-1)^\delta \, e^{\pm i \, Z_\lambda(\lambda)}},
\end{equation}
furthermore $h_j$ and $c_j$ denote the positions of holes and complex Bethe-roots respectively.
 $\eta$ is a 
small positive contour-integral parameter 
which should satisfy the inequalities:
\begin{equation}\label{ineqlemma}
0<\eta<\text{min} \{|\text{Im} \, p^{\pm}_\lambda| \}, \qquad |\text{Im}\,\mu \pm \eta|<|\text{Im} \, p^{(f)}|,
\end{equation}
where $p^{\pm}_\lambda$ denotes those complex{\footnote{I.e. not real.}} poles of ${\cal F}^{(\lambda)}_{\pm}(\lambda)$, which are 
located the closest to the real axis.}

The summation formula (\ref{sumrule}) can be extended to the $|\text{Im}\,\mu|>|\text{Im} \, p^{(f)}|$ 
domain by an analytical continuation procedure being similar to the analytical continuation of the DDV equation to the whole complex plane \cite{ddv97}.

Using the Ansatz (\ref{SXa}) and the formula (\ref{Phiab}) together with the parameterizations 
(\ref{2.11}), (\ref{inho}),
 the linear equations (\ref{Sa}) take the form:
\begin{equation} \label{lineq1}
-i \, Z'(\lambda_a) \, X(\lambda_a)-2 \pi \, i \sum\limits_{b=1}^m K(\lambda_a-\lambda_b|\gamma) \, X(\lambda_b)=2 \pi \, i \, K(\lambda_a-\rho_n|\tfrac{\gamma}{2}), \quad a=1,..,m,
\end{equation}
where $\rho_n=+\rho$ if $n$ is even and $\rho_n=-\rho$ otherwise with $\rho$ given by (\ref{rho}). 
We transform (\ref{lineq1}) into integral equations with the help of (\ref{sumrule}). Since in (\ref{sumrule}) the integrand always contains a factor $Z'_\lambda(\lambda)$, 
it is convenient to parameterize the function $X(\lambda)$ as:
\begin{equation} \label{XG}
X(\lambda)=\frac{{\cal G}(\lambda)}{Z'_\lambda(\lambda)}.
\end{equation}
Then using (\ref{sumrule}), for the case of pure hole states, the linear equations can be rewritten in the form of the following linear set of integral equations:
\begin{equation} \label{llie1}
\begin{split}
Z'_\lambda(\lambda) \, {\cal G}(\lambda)&+\! \!\! \int\limits_{-\infty}^{\infty}  d\lambda' \, K(\lambda-\lambda'|\gamma) \, {\cal G}(\lambda') - \\
&-\! \! \sum\limits_{\alpha=\pm} \int\limits_{-\infty}^{\infty} d\lambda' \,  K(\lambda-\lambda'+i \, \alpha \, \eta|\gamma) \, {\cal G}(\lambda'+i \, \alpha \, \eta) \, {\cal F}^{(\lambda)}_{\alpha}(\lambda'+i \, \alpha \, \eta)=\\
&=-2 \pi \, K(\lambda-\rho_n|\tfrac{\gamma}{2})+\sum\limits_{j=1}^{m_H} \, K(\lambda-h_j|\gamma) \, X_j,
\end{split}
\end{equation}
where
\begin{equation} \label{Xj}
X_j=X(h_j), \qquad j=1,...,m_H,
\end{equation}
such that they should satisfy the discrete set of equations:
\begin{equation} \label{eqXj}
X_j=\frac{{\cal G}(h_j)}{Z'_\lambda(h_j)}, \qquad j=1,..., m_H.
\end{equation}
Similarly, (\ref{sumrule}) allows us to rephrase (\ref{e_3}) as:
\begin{equation} \label{en1}
\begin{split}
\langle e_n \rangle_\lambda=\sum\limits_{j=1}^{m_H} \, X_j-\int\limits_{-\infty}^{\infty} \frac{d\lambda}{2 \pi} \, {\cal G}(\lambda)+
\sum\limits_{\alpha=\pm} \int\limits_{-\infty}^{\infty} \frac{d\lambda}{2 \pi} \,{\cal G}(\lambda+i \, \alpha \, \eta) \, {\cal F}^{(\lambda)}_{\alpha}(\lambda+i \, \alpha \, \eta).
\end{split}
\end{equation}
Acting{\footnote{Appendix \ref{appA}. contains some Fourier-transforms, which are necessary to do these
computations.}} $(1+K)^{-1}$ on (\ref{llie1}), the equations take the form:
\begin{equation} \label{llie2}
\begin{split}
 {\cal G}(\lambda)&- \sum\limits_{\alpha=\pm} \int\limits_{-\infty}^{\infty} d\lambda' \,  G_{\lambda}(\lambda-\lambda'+i \, \alpha \, \eta) \, {\cal G}(\lambda'+i \, \alpha \, \eta) \, {\cal F}^{(\lambda)}_{\alpha}(\lambda'+i \, \alpha \, \eta)=\\
&=-\frac{\pi}{\gamma} \, \frac{1}{\cosh\left(\tfrac{\pi}{\gamma}(\lambda-\rho_n)\right)}+\sum\limits_{j=1}^{m_H} \, G_\lambda(\lambda-h_j) \, X_j,
\end{split}
\end{equation}
where $G_\lambda(\lambda)$ is related to the kernel of DDV equation (\ref{G}) by:
\begin{equation} \label{Gl}
G_\lambda(\lambda)=\frac{1}{2 \gamma} \, G\left(\tfrac{\pi}{\gamma} \lambda \right), \quad \text{with} \quad \gamma=\tfrac{\pi}{p+1}.
\end{equation}
With the help of the integrated form of (\ref{llie2}), the term $\int\limits_{-\infty}^{\infty} \frac{d\lambda}{2 \pi} \,{\cal G}(\lambda)$ can be eliminated from (\ref{en1}).
Finally, one ends up with the following formula for the expectation value: 
\begin{equation} \label{enlatt}
\begin{split}
\langle e_n \rangle_\lambda & =\tfrac{1}{2}-\tfrac{1}{2} \langle \sigma^z_n \rangle_\lambda, \\
\tfrac{1}{2} \langle \sigma^z_n \rangle_\lambda &=- \frac{1}{2(1-\tfrac{\gamma}{\pi})} \, \left\{ 
\sum\limits_{j=1}^{m_H} \, X_j+
\sum\limits_{\alpha=\pm} \int\limits_{-\infty}^{\infty} \frac{d\lambda}{2 \pi} \,{\cal G}(\lambda+i \, \alpha \, \eta) \, {\cal F}^{(\lambda)}_{\alpha}(\lambda+i \, \alpha \, \eta)
\right\}.
\end{split}
\end{equation}
Equations (\ref{llie2}), (\ref{eqXj}) and (\ref{enlatt}) constitutes our final lattice results, 
which serve as the starting point to compute the continuum limit of the expectation values of the $U(1)$ 
current of the MT/SG theories.

\section{The continuum limit}
 
 The continuum limit is the appropriate $N \to \infty$ of our equations. Using (\ref{rho}), the equations (\ref{eqXj}),
  (\ref{llie2}) and (\ref{enlatt}) can be expanded at large $N$ in a series 
 of $1 \over N$, such that the leading power is $\tfrac{1}{N}$. From (\ref{j0na}) and (\ref{j1na}) one can see, that 
 the continuum result will be proportional to this leading order coefficient:
\begin{equation} \label{GXcont}
 {\cal G}^{cont}(\lambda)\sim\lim\limits_{N \to \infty} \, N \, {\cal G}(\lambda), \qquad X_j^{cont}\sim \lim\limits_{N \to \infty} \, N \, X_j.
\end{equation}
All equations we have for the computation of the expectation value of $\sigma^z_n$ are linear. 
This is why using (\ref{j0na}) and (\ref{j1na}), one can take their appropriate 
linear combinations to get the continuum 
expressions corresponding to the components of the $U(1)$ current.
The equations in the continuum limit and in rapidity convention take the form:
\begin{equation} \label{lliecont}
\begin{split}
{\cal G}^{(\mu)}(\theta)&-\sum\limits_{\alpha =\pm} \int\limits_{-\infty}^{\infty} \! \frac{d\theta'}{2 \pi} G(\theta-\theta'+i \, \alpha \, \eta)
{\cal G}^{(\mu)}(\theta'+i \, \alpha \, \eta) \, {\cal F}_{\alpha} (\theta'+i \, \alpha \, \eta)
=\\
&=-{\cal K}_\mu(\theta)+\sum\limits_{j=1}^{m_H} \, G(\theta-H_j) \, X_j^{(\mu)}, \\
X_j^{(\mu)}&=\frac{{\cal G}^{(\mu)}(H_j)}{Z'(H_j)}, \qquad j=1,...,m_H, \quad \mu=0,1. \\
\end{split}
\end{equation}
\begin{equation} \label{jmu}
\langle j_\mu(x) \rangle_H=-\frac{p+1}{ p} \, \left\{ \sum\limits_{j=1}^{m_H}  X_j^{(\mu)} +
\sum\limits_{\alpha =\pm} \int\limits_{-\infty}^{\infty} \! \frac{d\theta'}{2 \pi} 
{\cal G}^{(\mu)}(\theta'+i \, \alpha \, \eta) \, {\cal F}_{\alpha} (\theta'+i \, \alpha \, \eta)\right\},
\end{equation}
where
\begin{equation} \label{calF}
{\cal F}_{\pm}(\theta)=\frac{(-1)^\delta \, e^{\pm i \, Z(\theta)}}
{1+(-1)^\delta \, e^{\pm i \, Z(\theta)}}, 
\end{equation}
and the operator dependent source term reads as:
\begin{equation} \label{Kmu}
{\cal K}_\mu(\theta)=\left\{ 
\begin{array}{c}
{\cal M} \, \cosh(\theta), \qquad \mu=0, \\
{\cal M} \, \sinh(\theta), \qquad \mu=1.
\end{array}\right.
\end{equation}
Here the index $\mu=0,1$ corresponds to the lower index of the current $J_\mu$,
 $G(\theta)$ is the kernel (\ref{G}) of the DDV equation, and $\eta$ is a small
positive contour integral parameter which must satisfy the inequalities:
\begin{equation} \label{etacont}
0<\eta<\text{min}\{|\text{Im} \, p^{(\pm)}_j|\},  \qquad |\text{Im} \, \theta \pm \eta|<\text{min}(1,p) \, \pi,
\end{equation}
where $p^{(\pm)}_j$ denotes those poles of ${\cal F}_{\pm}(\theta)$ which are not real.

At the notation of the expectation value, we denoted that in the continuum limit we think of the state as if 
it was characterized by the holes.
For completeness we give how the continuum quantities of (\ref{lliecont}) are related to those of the lattice:
\begin{equation} \label{GXrel}
\begin{split}
{\cal G}^{(0)}(\theta)=\lim\limits_{N \to \infty} \tfrac{N}{2 L} \left\{\tfrac{\gamma}{\pi}{\cal{G}}^{(e_{2n})}(\tfrac{\gamma}{\pi}\theta)
+\tfrac{\gamma}{\pi}{\cal{G}}^{(e_{2n-1})}(\tfrac{\gamma}{\pi}\theta)\right\}, \\
{\cal G}^{(1)}(\theta)=\lim\limits_{N \to \infty} \tfrac{N}{2 L} \left\{\tfrac{\gamma}{\pi}{\cal{G}}^{(e_{2n})}(\tfrac{\gamma}{\pi}\theta)
-\tfrac{\gamma}{\pi}{\cal{G}}^{(e_{2n-1})}(\tfrac{\gamma}{\pi}\theta)\right\},
\end{split}
\end{equation}
where at the right hand side we indicated as an upper index the lattice operator, the unknown of whose linear 
problem{\footnote{Here by linear problem we mean, the linear problem which enables one to compute the expectation value. Namely, the set of equations:
(\ref{eqXj}) and (\ref{llie2}), (\ref{enlatt})}} should be considered. Finally we note that formula (\ref{jmu}) for $\langle j_\mu(x) \rangle_H$ is not the 
final answer to the expectation value of the $U(1)$ current. In the next subsection at the investigation of the charge 
density, it will turn out that $\langle j_\mu(x) \rangle_H$
 is still not the real expectation value of the $U(1)$ current in the quantum field theory, but it should be 
modified with an appropriate renormalization factor.

\subsection{The solution of equations}\label{solutions}

In this section we relate the equations (\ref{lliecont}) describing the expectation 
values of the $U(1)$ current to the counting-function
 of the DDV-equation (\ref{DDVcont}) corresponding to the sandwiching state. 
Indeed it turns out that the solutions of (\ref{lliecont})
are related to certain derivatives of $Z(\theta)$. The solutions we get, imply 
the relation  (\ref{Jjmu}).
 
 \subsubsection{The charge density case}

Let us start with the $\mu=0$ case, which corresponds to the expectation value of the charge density. Comparing (\ref{lliecont}) 
with the derivative of the DDV equation (\ref{DDVcont}) with respect to $\theta$, it turns out that the solution of $(\ref{lliecont})$
can be expressed as:
\begin{equation} \label{calG0}
\begin{split}
{\cal G}^{(0)}(\theta)&=-\tfrac{1}{L} Z'(\theta), \\
X^{(0)}_j&=-\tfrac{1}{L}, \quad j=1,..,m_H, 
\end{split}
\end{equation}
and the expectation value between $m_H$ solitons is given by the the formula:
\begin{equation} \label{j0exp}
\langle j_0(x) \rangle_H=\tfrac{p+1}{p} \tfrac{m_H}{L}.
\end{equation}
This formula requires some explanation.
In the quantum field theory each soliton carries topological charge $Q=+1$. The expectation value of the topological charge in an $m_H$ soliton state 
is $\langle Q \rangle_H=m_H$. Since the charge is the integral of the charge density operator, whose expectation value has no space-time dependence, 
in the continuum theory the expectation value of the charge density should be $\tfrac{m_H}{L}.$ It can be seen, that the result in (\ref{j0exp}) agrees with the
 expected one apart from a global coupling dependent factor of $\tfrac{p+1}{p}.$ We got (\ref{j0exp}) by  
identifying the topological charge of the continuum field theory with twice the z-component of the spin of the lattice 
model, $Q\sim 2 \, S_z$. In view of the quantum field theory interpretation, formula (\ref{j0exp}) suggests that
 instead of (\ref{j0na}) and (\ref{j1na})
 the correct identification between the lattice and continuum operators is given by the formulas:
 \begin{equation} \label{Jmuop}
\begin{split}
J_0(x)|_{x=na}=\tfrac{ p}{p+1} \, j_0(x)|_{x=na} = \frac{p}{p+1}   \frac{N}{L}\frac{\sigma^z_{2n}+\sigma^z_{2n-1}}{2}, \\
J_1(x)|_{x=na}=\tfrac{ p}{p+1} \, j_1(x)|_{x=na} = \frac{p}{p+1}   \frac{N}{L}\frac{\sigma^z_{2n}-\sigma^z_{2n-1}}{2}.
\end{split}
\end{equation}
Consequently, we conclude that $\langle j_\mu(x) \rangle_H$ given in (\ref{lliecont}) is not the final answer in the quantum field theory (QFT), because it has to be modified 
by the renormalization factor $Z_p=\frac{p}{p+1}.$ Thus, the real QFT result is given by:
 \begin{equation} \label{Jmu}
\begin{split}
\langle J_\mu(x) \rangle_H&=-\sum\limits_{j=1}^{m_H}  X_j^{(\mu)} -
\sum\limits_{\alpha =\pm} \int\limits_{-\infty}^{\infty} \! \frac{d\theta'}{2 \pi} 
{\cal G}^{(\mu)}(\theta'+i \, \alpha \, \eta) \, {\cal F}_{\alpha} (\theta'+i \, \alpha \, \eta).
\end{split}
\end{equation}
To summarize: (\ref{lliecont}) and (\ref{Jmu}) constitutes our final equations for computing the diagonal matrix
 elements of $J_\mu(x).$ 
 
 \subsubsection{The case of $J_1(x)$}
 
 The equations (\ref{lliecont}) also for $\mu=1$ are  related to a certain derivative of the counting-function. 
The counting-function depends on the spectral parameter $\theta$, on $\ell={\cal M} L$ the dimensionless length
 of the system and on the hole positions{\footnote{Specifying the state, the quantum numbers of holes 
in the continuum version of (\ref{QHk}) are fixed. 
I.e. They are $\ell$ independent.}}, which are also $\ell$ dependent.
 Then, differentiating (\ref{DDVcont}) with respect to $\ell$, one can recognize that
 ${\cal G}^{(1)}(\theta)$ of (\ref{lliecont}) is related 
 to the $\ell$-derivative of the counting-function as follows:
 \begin{equation} \label{calG1}
\begin{split}
{\cal G}^{(1)}(\theta)&=-{\cal M}\, \frac{d}{d \ell} \, Z(\theta|\vec{H}(\ell),\ell), \\
X^{(1)}_j&=-{\cal M} \, H'_j(\ell), \qquad j=1,.., m_H,
\end{split}
\end{equation}
where we explicitly wrote out the $H_j$ and $\ell$ dependence of $Z(\theta).$ 
With the help of (\ref{calG1}), one can show that $\langle J_1(x) \rangle_H$ can also be rephrased
 as the $\ell$-derivative of a quantity, which can be expressed directly in terms of the solution of the DDV equations;
\begin{equation} \label{J1der}
\begin{split}
\langle J_1(x) \rangle_H&={\cal M} \, \frac{d}{d \ell} \Lambda_{1}(\ell), \\
\Lambda_{1}(\ell)&=\sum\limits_{j=1}^{m_H} H_j(\ell)- \int\limits_{-\infty}^{\infty} \frac{d \theta}{2 \, \pi\, i} \, \left\{ L_{+}(\theta+i \, \eta)-L_{-}(\theta-i \, \eta)\right\}.
\end{split}
\end{equation}

We note, that in (\ref{J1der}) $L_+$ and $L_-$ have different signs under the integration. This fact has a remarkable 
consequence concerning the TBA description of this expectation value. Namely, if one considers the TBA description
 \cite{FW2,BHsg} of 
the model at the points $1<p\in \mathbb{Z}_+$, where the system is described by a ${\cal D}_{p+1}$-type TBA-system, 
then it becomes obvious, that (\ref{J1der}) cannot be expressed in terms of the Y-functions corresponding to the 
massive TBA node{\footnote{This is so, because the relation between $L_{\pm}$ and $Y_1$ the massive Y-function is given by \cite{BHsg}: $L_+(\theta+i \tfrac{\pi}{2})+L_-(\theta-i \tfrac{\pi}{2})=\ln(1+Y_1(\theta)).$}}. This implies, that the TBA conjectures \cite{LM99,saleur,Pozsg11,Pozsg13,PST14} for purely 
elastic scattering theories, cannot be valid in this non-diagonally scattering theory. Earlier a similar conclusion 
has been drawn in \cite{BucT}.

\section{The large volume expansion} 
 
In this section we solve our equations (\ref{lliecont}) in the context of a systematic large volume expansion. 
The actual form of the representation we get, is very similar to those conjectured for purely elastic scattering theories \cite{LM99,saleur,Pozsg11,Pozsg13,PST14}. 
Nevertheless, since our model is not a diagonally scattering theory, 
our large volume series  differs from these TBA conjectures. 
 
In this section we will strongly rely on the method described in \cite{PST14} for the computation of the 
diagonal matrix elements of the trace of the stress-energy tensor in purely elastic scattering theories.
We can do this, because the DDV equation (\ref{DDVcont}) is formally similar to the TBA-equations 
of a purely elastic scattering theory containing two types of particles. To clarify this analogy better,
 as a first step we reformulate the DDV equation as a two-component TBA equation.  
(\ref{DDVcont}) contains $Z(\theta)$ along three different
 lines; along the real line and on the lines $\theta\pm i \, \eta$ with $\theta \in \mathbb{R}.$ 
When solving the equations, one has to compute $Z(\theta)$ on all these 3 lines.
To get a closed set of equations, we have to consider (\ref{DDVcont}) with left-hand sides $Z(\theta\pm i \, \eta)$ 
as well.
The two equations for $Z(\theta\pm i \, \eta)$ formally look like a two component TBA equation of a diagonally 
scattering theory. Let $\varepsilon_{\pm}(\theta)=Z(\theta\pm i \, \eta)$ and 
${\cal L}_{\pm}(\theta)=\ln(1+(-1)^\delta \, e^{\pm i\, \varepsilon_{\pm}(\theta) })$, then the TBA-like form of 
(\ref{DDVcont}) reads as:
\begin{equation} \label{DDVtba}
\begin{split}
\varepsilon_{\alpha}(\theta)={\cal S}_\alpha(\theta)+\sum\limits_{\beta=\pm}
\int\limits_{-\infty}^{\infty} \! \frac{d \theta'}{2 \pi} \,
\varphi_{\alpha \beta}(\theta-\theta') \, {\cal L}_\beta(\theta'), \quad \alpha=\pm,
\end{split}
\end{equation}
where ${\cal S}_\alpha(\theta)$ is the source term:
\begin{equation} \label{calS}
\begin{split}
{\cal S}_\alpha(\theta)=\ell \sinh (\theta+i \, \alpha \, \eta) 
+\sum\limits_{k=1}^{m_H} \, \chi(\theta+i \, \alpha \, \eta -H_k), 
\end{split}
\end{equation}
and 
\begin{equation} \label{tbamag}
\varphi_{\alpha \beta}(\theta)=i \, G(\theta+i \, (\alpha-\beta) \, \eta), \quad \alpha,\beta=\pm,
 \end{equation}
is  a symmetric matrix kernel. From the point of view of our later computations, the fact that the different 
quantities in (\ref{DDVtba}) are complex, does not matter. The only important property is that
 the kernel (\ref{tbamag}) is symmetric, i.e. $\varphi_{\alpha \beta}(\theta)=\varphi_{\beta \alpha}(-\theta).$ 

In a completely analogous way the linear equations (\ref{lliecont}) can also be rephrased by considering them along 
the lines $\theta\pm i\, \eta.$ In (\ref{lliecont}) the left-hand side describes the action of the linear operator 
on the unknown 
function and the right-hand side is the source term of the linear problem. 
Since these equations are linear, it is worth to consider the solutions of (\ref{lliecont})
 with different "elementary" source terms, from which the solution of the physical problem can be obtained by 
linear combinations.  
Consider in general the linear problems:
\begin{equation} \label{lliegen}
\begin{split}
{\cal G}^{[\alpha]}_A(\theta)-\sum\limits_{\beta =\pm} \int\limits_{-\infty}^{\infty} \! \frac{d\theta'}{2 \pi} \psi_{\alpha \beta}(\theta-\theta') \, {\cal G}^{[\beta]}_A(\theta') \,{\cal F}_{\beta}^{[\beta]} (\theta')=f_A^{[\alpha]}(\theta), \qquad \alpha=\pm,
\end{split}
\end{equation}
where for any function $f(\theta)$ we introduced the notation: $f^{[\pm]}(\theta)=f(\theta\pm i \, \eta),$
and $\psi_{\alpha \beta}(\theta)=\tfrac{1}{i} \varphi_{\alpha \beta}(\theta)$ is a symmetric kernel.
In (\ref{lliegen}) $f_A^{[\alpha]}(\theta)$ denotes the elementary source term indexed by $A.$

If the argument of the "elementary" solution is not shifted, we denote it simply ${\cal G}_A(\theta)$
 and it satisfies the equations:
\begin{equation} \label{lliegen0}
\begin{split}
{\cal G}_A(\theta)-\sum\limits_{\beta =\pm} \int\limits_{-\infty}^{\infty} \! \frac{d\theta'}{2 \pi} \, \psi_{\alpha \beta}(\theta-\! \theta'\!-i \, \alpha \, \eta) \, {\cal G}^{[\beta]}_A(\theta') \, {\cal F}_{\beta}^{[\beta]} (\theta')=f_A (\theta), \qquad \alpha=\pm.
\end{split}
\end{equation}
The elementary solutions from which the physical solutions of (\ref{lliecont}) and (\ref{Jmu}) can be combined are
characterized by their source terms in (\ref{lliegen0}) and they are as follows:
\begin{eqnarray} \label{GAlist}
{\cal G}_{{\cal K}_\mu}(\theta) \quad &\leftrightarrow & \quad  f_{{\cal K}_\mu}(\theta)={\cal K}_\mu(\theta), \qquad \qquad 
\quad \mu=0,1. \\
{\cal G}_{j}(\theta) \quad &\leftrightarrow & \quad  f_j(\theta)=-G(\theta-H_j), \qquad j=1,...,m_H, \label{GAj}\\
{\cal G}_{u}(\theta) \quad &\leftrightarrow & \quad  f_u(\theta)=1. \label{GAlast}
\end{eqnarray}
As a consequence of equations (\ref{lliegen}), for any pair of indexes the following identities hold:
 \begin{equation} \label{idAB}
\begin{split}
\sum\limits_{\alpha=\pm} \int\limits_{-\infty}^{\infty} \frac{d \theta}{2 \pi} \, f_A^{[\alpha]}(\theta) \,
{\cal G}^{[\alpha]}_B(\theta) \, {\cal F}_{\alpha}^{[\alpha]}(\theta)=
\sum\limits_{\alpha=\pm} \int\limits_{-\infty}^{\infty} \frac{d \theta}{2 \pi} \, f_B^{[\alpha]}(\theta) \,
{\cal G}^{[\alpha]}_A(\theta) \, {\cal F}_{\alpha}^{[\alpha]}(\theta).
\end{split}
\end{equation}

Now we show how the exact Gaudin-matrix enters the large volume expansion and how 
one can express the solutions of (\ref{lliecont}) in terms of the elementary solutions (\ref{GAlist})-(\ref{GAlast}).
First, we consider the integral equation in (\ref{lliecont}) as if $X^{(\mu)}_j$ were arbitrary parameters. Then
 using (\ref{lliegen}) and (\ref{GAlist})-(\ref{GAlast}) the solution can be written as:
\begin{equation} \label{GmuX}
\begin{split}
{\cal G}^{(\mu)}(\theta)=-{\cal G}_{{\cal K}_\mu}(\theta)-\sum\limits_{j=1}^{m_H} {\cal G}_j(\theta) \, X^{(\mu)}_j.
\end{split}
\end{equation}
However we know from (\ref{lliecont}) that $X^{(\mu)}_j$s are not independent from ${\cal G}^{(\mu)}(\theta),$ 
but they are related by: ${{\cal G}^{(\mu)}(H_j)=X^{(\mu)}_j \, Z'(H_j)}.$ Inserting this relation into 
(\ref{GmuX}) taken at $\theta=H_k,$ one ends up with the discrete set of equations for $X^{(\mu)}_j$ as follows:
\begin{equation} \label{Xeq1}
\begin{split}
\sum\limits_{j=1}^{m_H} \left\{ Z'(H_k) \, \delta_{jk}+{\cal G}_j(H_k) \right\}\, X^{(\mu)}_j
=-{\cal G}_{{\cal K}_\mu}(H_k), \qquad k=1,...,m_H.
\end{split}
\end{equation}
From (\ref{DDVcont}) and (\ref{QHk}), it follows that the matrix entering (\ref{Xeq1}) is nothing, but the 
Gaudin-matrix of physical excitations over the 
Dirac-see, which we call exact Gaudin-matrix:
\begin{equation} \label{exactGaudin}
\begin{split}
\hat{\Phi}_{kj}(\vec{H})=\frac{d}{d H_j} Z(H_k|\vec{H})=Z'(H_k) \, \delta_{jk}+{\cal G}_j(H_k), \qquad j,k=1,..,m_H.
\end{split}
\end{equation}
Using (\ref{GmuX}), (\ref{Xeq1}) and (\ref{exactGaudin}) finally we get:
\begin{equation} \label{Xr1}
X^{(\mu)}_k=-\sum\limits_{j=1}^{m_H} \hat{\Phi}^{-1}_{kj}(\vec{H})\,{\cal G}_{{\cal K}_\mu}(H_j), \qquad k=1,..,m_H.
\end{equation}
\begin{equation} \label{Gr1}
\begin{split}
{\cal G}^{(\mu)}(\theta)=-{\cal G}_{{\cal K}_\mu}(\theta)
+\sum\limits_{k=1}^{m_H}\sum\limits_{j=1}^{m_H} {\cal G}_k(\theta) \, \hat{\Phi}_{kj}^{-1}(\vec{H})\,{\cal G}_{{\cal K}\mu}(H_j).
\end{split}
\end{equation}
The last missing piece is the expression of $\langle J_\mu(x) \rangle_H$ in terms of the "elementary" solutions.
This can be computed by inserting (\ref{Xr1}) and (\ref{Gr1}) into (\ref{Jmu}) and by using the identity:
\begin{equation} \label{idu}
\begin{split}
\sum\limits_{\alpha=\pm} \int\limits_{-\infty}^{\infty} \frac{d \theta}{2 \pi} \,
{\cal G}^{[\alpha]}_j(\theta) \, {\cal F}_{\alpha}^{[\alpha]}(\theta)=1-{\cal G}_u(H_j), \quad j=1,..,m_H,
\end{split}
\end{equation}
which can be derived by using (\ref{idAB}).
The final result is as follows:
\begin{equation} \label{Jr1}
\begin{split}
\langle J_\mu(x) \rangle_H=\sum\limits_{\alpha=\pm} \int\limits_{-\infty}^{\infty} \frac{d \theta}{2 \pi} \,
{\cal G}^{[\alpha]}_{{\cal K}_\mu}(\theta) \, {\cal F}_{\alpha}^{[\alpha]}(\theta)+\sum\limits_{k=1}^{m_H} 
\sum\limits_{j=1}^{m_H} {\cal G}_u(H_j) \, \hat{\Phi}^{-1}_{jk}(\vec{H}) \, {\cal G}_{{\cal K}_\mu}(H_k).
\end{split}
\end{equation}
The first term in (\ref{Jr1}) corresponds to the so-called vacuum contribution \cite{Pozsg13,PST14}.
Constructing the all order large volume solution of (\ref{lliegen}) for $A={\cal K}_\mu$,
 it can be written as an infinite series similar to that of LeClair and Mussardo \cite{LM99,saleur,Pozsg11}. 
Performing carefully the calculations one obtains for the vacuum piece the result as follows:
\begin{equation} \label{Jmuvac}
\begin{split}
\langle J_\mu(x) \rangle_H\big|_{vac}=\sum\limits_{n_+=0}^\infty \sum\limits_{n_-=0}^\infty \,
\frac{1}{n_+! \, n_-!} \, \int \prod\limits_{i=1}^{n_++n_-} \frac{d \theta_i}{2 \pi} \,
\prod\limits_{i=1}^{n_+} {\cal F}_+(\theta+i \, \eta) \, \prod\limits_{i=n_++1}^{n_++n_-} {\cal F}_-(\theta-i \, \eta)\\ \times
F^{J_\mu}_c(\theta_1\!+\!i \, \eta,...,\theta_{n_+}\!+\!i \, \eta,\theta_{n_++1}\!-\!i \, \eta,...,\theta_{n_++n_-}\!-\!i \, \eta),
\end{split}
\end{equation}
where $F^{J_\mu}_c$ denotes the connected diagonal form factors of the operator $J_\mu(x)$ between pure soliton states. 
Since $J_\mu$ is a conserved current, its connected form-factors can be determined by simple modifications of the arguments
of references \cite{LM99} and \cite{saleur}. 
The explicit form of $F^{J_\mu}_c$ is given by the compact formula{\footnote{In (\ref{Fc}) 
the $\langle \theta|\theta' \rangle=2 \pi \, \delta(\theta-\theta')$ normalization for the continuum 
 states is assumed.}}:  
\begin{equation} \label{Fc}
\begin{split}
F^{J_\mu}_c(\theta_1,\theta_2,...,\theta_n)=\sum\limits_{\sigma\in S_n} {\cal K}_\mu(\theta_{\sigma(n)}) \,
\prod\limits_{j=1}^{n-1} G(\theta_{\sigma(j)}-\theta_{\sigma(j+1)}), 
\end{split}
\end{equation}
where $\sigma$ denotes the elements of the the symmetric group $S_n.$ 

Before turning to the second term is the rhs. of (\ref{Jr1}) it is worth to recall the conjecture of  \cite{Pozsg13,PST14} 
for the diagonal matrix elements of local operators in purely elastic scattering theories. 

{\it The conjecture for purely elastic scattering theories states, that the exact finite volume expectation value 
of any local operator ${\cal O}(x)$ can be written as:
\begin{equation} \label{Ps1}
\begin{split}
\langle H_1,...,H_n|{\cal O}(x)|H_1,...,H_n \rangle=&\frac{1}{\rho(H_1,..,H_n)} \\
& \times \sum\limits_{\{H_+\}\cup \{H_-\}} {\cal D}^{{\cal O}}(\{H_+\}) \, \rho(\{H_-\}|\{H_+\}),
\end{split}
\end{equation}
where $\rho(\vec{H})$ is the determinant of the exact Gaudin-matrix:
\begin{equation} \label{ro}
\rho(H_1,..,H_n)=\text{det} \, \hat{\Phi}(\vec{H}),
\end{equation}
the sum in (\ref{Ps1}) runs for all bipartite partitions of the rapidities of the sandwiching state:
$\{H_1,..,H_n \}=\{H_+\}\cup\{H_-\},$ such that
\begin{equation} \label{ropm}
\rho(\{H_+\}|\{H_-\}=\text{det} \, \hat{\Phi}_+(\vec{H}),
\end{equation}
with $\hat{\Phi}_+(\vec{H})$ being the submatrix of $\hat{\Phi}(\vec{H})$ corresponding to the subset $\{H_+\}.$
The most important part in (\ref{Ps1}) is the form of the so-called dressed-form factor ${\cal D}^{\cal O}(\{H_+\}).$ 
It is expressed as an infinite sum in terms of the connected diagonal form-factors of the theory:
\begin{equation} \label{PD1}
\begin{split}
{\cal D}^{\cal O}(\{H_1,...,H_l\})=&\sum\limits_{n_1,..,n_k}^\infty \frac{1}{\prod\limits_i \, n_i!} 
 \int\limits_{-\infty}^{\infty} \prod\limits_{j=1}^{\scriptsize{\sum_i n_i}} 
 \frac{d \theta_j}{2 \pi \, \left[ 1+e^{\varepsilon_{\beta_j}(\theta_j)} \right]}\, \\
&\times F^{\cal O}_{2l,2 n_1,..,2 n_k}(H_1,..,H_l,\theta_1,...,\theta_{\tiny{\sum_i n_i}}),
\end{split}
\end{equation}
where $\varepsilon_{\beta_j}(\theta_j)$ is the pseudoenergy of the particle of type $\beta_j$ in the 
TBA equations of the model and $ F^{\cal O}_{2l,2 n_1,..,2 n_k}$ is the connected diagonal form 
factor of the operator ${\cal O}$ in the theory, such that $n_i$ denotes the number of particles of type 
$\beta_i$ in the set $\{\theta_1,..,\theta_{\tiny{\sum_i n_i}}\}.$}

Now we can turn to compute the second term in the rhs. of (\ref{Jr1}). In this paper we consider the 
expectation values of $J_\mu$ between pure soliton states. In the SG/MT model solitons scatter diagonally 
among themselves. In this respect the pure soliton sector is very similar to a purely elastic scattering 
theory. Consequently, we expect a final result similar to the conjecture (\ref{Ps1}) for the soliton expectation 
values of $J_\mu.$ This is why we will show that the second term in the right-hand side of (\ref{Jr1}) can be brought 
into the form of (\ref{Ps1}) such that our dressed form factors are defined as the coefficients of 
$\rho(\{H_-\}|\{H_+\}) \over \rho(H_1,...,H_n)$ in the sum. The expression we want to bring into the 
form of (\ref{Ps1}) reads as follows{\footnote{The term corresponding to $\{H_+\}=\emptyset$ is given by the
vacuum contribution (\ref{Jmuvac}). }}:
\begin{equation} \label{Jrex}
\begin{split}
\langle J_\mu(x) \rangle_H\big|_{ex}=\sum\limits_{k=1}^{m_H} 
\sum\limits_{j=1}^{m_H} {\cal G}_u(H_j) \, \hat{\Phi}^{-1}_{jk}(\vec{H}) \, {\cal G}_{{\cal K}_\mu}(H_k).
\end{split}
\end{equation}
Our sandwiching state is composed of $m_H$-solitons. The key point in the computation is that the inverse 
Gaudin-matrix can be expanded in terms of its minors as follows \cite{PST14,PST50}.
\begin{equation} \label{GC}
\begin{split}
\hat{\Phi}_{ij}^{-1}=\frac{{\cal C}_{ij}}{\text{det} \, \hat{\Phi}}, \qquad i,j=1,..,m_H,
\end{split}
\end{equation}
where ${\cal C}_{ij}$ is the co-factor matrix. It is given by:
\begin{equation} \label{calC}
\begin{split}
{\cal C}_{ij}=\left\{\begin{array}{c}
\text{det} \, \hat{\Phi}(\{i\}), \qquad \qquad \qquad \qquad \qquad \qquad \qquad \qquad \qquad \qquad \quad i=j, \\
\sum\limits_{n=0}^{m_H-2} \, \sum\limits_{\{\alpha \}} \,
(-1)^{n+1} \, \hat{\Phi}_{i \alpha_1} \, \hat{\Phi}_{\alpha_1 \alpha_2} \dots \hat{\Phi}_{\alpha_n j} \,
\, \text{det} \, \hat{\Phi}(\{j,i,\alpha_1,...,\alpha_n \}), \quad i \neq j,
\end{array}\right.
\end{split}
\end{equation}
where $\{\alpha \}=\{1,2,...,m_H\} \setminus \{i,j\}$ and $\hat{\Phi}(\{ {\cal I} \})$ denotes the matrix 
obtained by omitting from $\hat{\Phi}$ the rows and columns indexed by the set $\{{\cal I}\}.$

First, one has to construct the all order large volume solution{\footnote{In the actual computations it is
convenient to write ${\cal G}_A(H_j) \to {\cal G}^{[\pm]}_A(H_j\mp i \, \eta)$ and iterate the two-component 
equations (\ref{lliegen}). }} of (\ref{lliegen}) 
for $A\in\{u,{\cal K}_0,{\cal K}_1 \}$ and to insert (\ref{GC}) with (\ref{calC}) into (\ref{Jrex}). 
Then after the careful bookkeeping of the terms being identical due to appropriate permutations
of the variables,  
one obtains the following expression for the dressed form factors between 
soliton states for $J_\mu$:
\begin{equation} \label{JDdress}
\begin{split}
{\cal D}^{J_\mu}(\{H_1,...,H_n\})=\sum\limits_{n_+=0}^\infty \sum\limits_{n_-=0}^\infty 
\frac{1}{n_+! \, n_-!} \! \int \! \prod\limits_{i=1}^{n_++n_-} \! \! \frac{d \theta_i}{2 \pi} \!
\prod\limits_{i=1}^{n_+} \! {\cal F}_+(\theta_i+i \, \eta) \! \! \! \prod\limits_{i=n_++1}^{n_++n_-} \! \!
{\cal F}_-(\theta_i-i \, \eta)\\
\times 
F^{J_\mu}_c(H_1,H_2,...,H_n,\theta_1\!+\!i \, \eta,...,\theta_{n_+}\!+\!i \, \eta,\theta_{n_++1}\!-\!i \, \eta,...,\theta_{n_++n_-}\!-\!i \, \eta).
\end{split}
\end{equation}
Then the expectation value of $J_\mu$ between pure soliton states is given by a formula being completely analogous
to (\ref{Ps1}):
\begin{equation} \label{Hs1}
\begin{split}
\langle H_1,...,H_{m_H}|{\cal J_\mu}(x)|H_1,...,H_{m_H} \rangle=&\frac{1}{\rho(H_1,..,H_{m_H})} \\
& \times \sum\limits_{\{H_+\}\cup \{H_-\}} {\cal D}^{{\cal J_\mu}}(\{H_+\}) \, \rho(\{H_-\}|\{H_+\}),
\end{split}
\end{equation}
with ${\cal D}^{{\cal J_\mu}}(\{H_+\})$ is given by (\ref{JDdress}). 
The result (\ref{JDdress}) requires some interpretation in view of previous results for purely elastic scattering 
theories \cite{Pozsg13,PST14}, which we summarized in (\ref{Ps1}). 
If one compares our results (\ref{Hs1}), (\ref{JDdress}) to the purely elastic TBA conjectures (\ref{Ps1}),(\ref{PD1}),
it is easy to recognize that the difference is present only in the actual form of the dressed form factors. 
Moreover at leading order in the volume, when the integral terms in (\ref{JDdress}) and (\ref{PD1}) can be neglected, 
in accordance with \cite{Palmai13}
 our formula agrees with the conjecture for purely elastic scattering theories \cite{PT08a,PT08b}. The reason for this 
might be, that pure soliton states form a purely elastic scattering subsector in the scattering theory of the SG/MT 
model. On the other hand, if one takes a look at the exponentially small in volume corrections, which are given by 
the integral terms in (\ref{JDdress}) and (\ref{PD1}), it becomes obvious that (\ref{PD1}) cannot 
describe{\footnote{The same fact was recognized in \cite{BucT}.}}
the SG/MT model by simply substituting the massive pseudoenergy of the TBA equations \cite{FW2,BHsg} of the SG-model into (\ref{PD1}).
This is because at the level of exponentially small in volume corrections, the  interactions 
between solitons and antisolitons will also contribute.
Apart from the differences between (\ref{JDdress}) and (\ref{PD1}) there is a remarkable similarity, too. Namely 
both formula contains the connected diagonal form factors of the operator sandwiched. 
Though we computed explicitely the diagonal matrix elements of only the components of the $U(1)$ current of the
SG/MT model, based on the remarkable similarity of our result with those obtained in purely elastic scattering theories \cite{Pozsg13,PST14}, 
we make the following conjecture:
\newline {\bf Conjecture:}
{\it
For any local operator ${\cal O}(x)$ in the SG/MT model the expectation value in an $n$-soliton state is given by
(\ref{Ps1}), such that the dressed form factors are given by the formula:
\begin{equation} \label{ODdress}
\begin{split}
{\cal D}^{\cal O}(\{H_1,...,H_n\})=\sum\limits_{n_+=0}^\infty \sum\limits_{n_-=0}^\infty 
\frac{1}{n_+! \, n_-!} \! \int \! \prod\limits_{i=1}^{n_++n_-} \! \! \frac{d \theta_i}{2 \pi} \!
\prod\limits_{i=1}^{n_+} \! {\cal F}_+(\theta_i+i \, \eta) \! \! \! \prod\limits_{i=n_++1}^{n_++n_-} \! \!
{\cal F}_-(\theta_i-i \, \eta)\\
\times 
F^{\cal O}_c(H_1,H_2,...,H_n,\theta_1\!+\!i \, \eta,...,\theta_{n_+}\!+\!i \, \eta,\theta_{n_++1}\!-\!i \, \eta,...,\theta_{n_++n_-}\!-\!i \, \eta).
\end{split}
\end{equation}
where $F^{{\cal O}}_{c}$ denotes the connected diagonal form factors of $O(x)$ in pure soliton states
 and ${\cal F}_\pm(\theta)$ are an appropriate nonlinear expressions (\ref{calF}) of the counting function 
of the continuum theory. 
}

The further analytical and numerical tests of our conjecture are left for future investigations.

\section{Summary and outlook}

In this paper we argued that, through the light-cone lattice approach, the QISM admits an appropriate framework for 
computing the finite volume form factors of Massive-Thirring/sine-Gordon theories. We demonstrated that the QISM 
works efficiently, when the diagonal matrix elements of local operators are computed. 

Our approach is similar to that of \cite{JMS_IV,JMS_V}, where the finite temperature one-point functions 
of  all local operators of the sine-Gordon model have been computed, which corresponds to finite volume vacuum 
expectation values in our language. The main difference between the two approaches is that, the authors of  
\cite{JMS_IV,JMS_V} work in a picture, when the compactified direction is time and the compactification length 
corresponds to the inverse temperature, while we work in the other possible channel, when the space is compactified. 
This allows us to consider form factors of operators between all possible excited states of the model. Consequently, 
our method allows one to extend the results of \cite{JMS_IV,JMS_V}
from vacuum expectation values to compute diagonal matrix elements of local operators of the Massive-Thirring/sine-Gordon models. To be more precise our approach works for operators, which are
composed of Fermi fields and their derivatives in the MT model and for their bosonized counterparts 
in the SG model.

Nevertheless, in this paper we considered only a simple operator, 
the $U(1)$ current of the theory and computed its diagonal matrix elements between pure soliton states. 
Our results are given by the formulas
(\ref{lliecont}) and (\ref{Jmu}). The computation of an expectation value consists of three steps: \newline
1. First one should solve the DDV equation (\ref{DDVcont}) for the sandwiching state. \newline
2. Then, one should solve the linear equations (\ref{lliecont}). \newline
3. Finally, the solution of (\ref{lliecont}) should be inserted into (\ref{Jmu}). \newline
The whole procedure can be written in the form of a systematic large volume expansion (\ref{Hs1}), (\ref{JDdress}), in which the diagonal connected 
form factors of $J_\mu$ arise. The remarkable similarity of the large volume series of $J_\mu$ to the large volume series conjectures for diagonal 
matrix elements of local operators in purely elastic scattering theories \cite{LM99,saleur,Pozsg11,Pozsg13,PST14} 
made us to conjecture, that formulas (\ref{Ps1}) and (\ref{ODdress}) describe the pure solitonic finite volume 
 expectation values of any local operators of the Massive-Thirring/sine-Gordon models.

Beyond the results of this paper a lot of interesting questions are still open. 
It would be important to test the conjecture (\ref{Hs1}) and (\ref{ODdress}) for other operators than $J_\mu.$ 
As it was demonstrated in \cite{FT,FPT11} the truncated conformal space approach could be an appropriate method for these investigations.
It would be also interesting to know how the large volume series formulas (\ref{Ps1}), (\ref{ODdress}) and  (\ref{Hs1}), (\ref{JDdress})
 should be modified, when expectation values between not pure soliton
states are considered. 
And finally the computation of non-diagonal finite volume form factors would be also of great importance.

Beyond the light-cone lattice approach of \cite{ddvlc}, in the literature 
 there exists another integrable lattice regularization{\footnote{This lattice regularization 
is based on a spin $-\tfrac{1}{2}$ spin-chain, while the light-cone lattice approach uses a spin $+\tfrac{1}{2}$ chain.}} for the sine-Gordon model \cite{book,latsg}. Though, in the framework 
of this approach local operators \cite{Oota,gmn12} and their form factors \cite{gmn12} have been computed on the lattice, the 
continuum results are still missing. It would be also very interesting to see,
 whether this approach also allows one to 
compute diagonal matrix elements of local operators of the continuum theory. 

\vspace{1cm}
{\tt Acknowledgments}

\noindent 
The author would like to thank Zolt\'an Bajnok and J\'anos Balog for useful discussions.
This work was supported by OTKA grant under K109312. The author also would like to thank
 the support of an MTA-Lend\"ulet Grant.

\appendix

\section{Conventions of Fourier transformation} \label{appA}

In this short appendix we summarize our conventions for Fourier-transformation and provide
the Fourier-transform of some functions we used in section \ref{detSa}.

Our convention for the Fourier-transform of a function $f$ is given by:
\begin{equation} \label{Ff}
\tilde{f}(\omega)=\int\limits_{-\infty}^{\infty} dx\, e^{i \omega  x} \, f(x).
\end{equation}
The inverse transformation reads as:
\begin{equation} \label{iFf}
f(x)=\int\limits_{-\infty}^{\infty} \frac{dx}{2 \pi} e^{-i \omega  x} \, \tilde{f}(\omega).
\end{equation}
The Fourier-transform of the convolution two functions $f$ and $g$ is given by the product of individual
 Fourier-transforms:
 \begin{equation} \label{Ffg}
\begin{split}
(f \star g)(x)=\int\limits_{-\infty}^{\infty} dy \, f(x-y)\, g(y), \qquad \widetilde{(f \star g)}(\omega)=\tilde{f}(\omega) \, \tilde{g}(\omega).
\end{split}
\end{equation}
When deriving the linear equations (\ref{llie2}) one needs the Fourier-transform of $K(\lambda|\gamma)$ of (\ref{K1}). It is given by the formula: 
\begin{equation} \label{Ktf}
\tilde{K}(\omega|\gamma)=\frac{\sinh\left[ \tfrac{\pi \omega}{2} \left(1-\tfrac{2 \gamma}{\pi}\right)\right]}{\sinh\left( \tfrac{\pi \omega}{2}\right)}.
\end{equation}
 The following inverse transform played important role at the determination of the source term in (\ref{llie2}): 
\begin{equation} \label{ich}
\begin{split}
\int\limits_{-\infty}^{\infty} \frac{dx}{2 \pi} e^{-i \omega  \lambda} \, \frac{\tilde{K}(\omega|\tfrac{\gamma}{2})}{1+\tilde{K}(\omega|\gamma)}=\frac{1}{2 \,\gamma}\,
\frac{1}{\cosh(\tfrac{\pi \lambda}{\gamma})}.
\end{split}
\end{equation}

\end{document}